\documentstyle[12pt,aaspp4,flushrt]{article}

\newcommand{\msun}{\mbox{ M$_{\odot}$}}
\newcommand{\bq}{\begin{equation}}
\newcommand{\eq}{\end{equation}}
\newcommand{\ergs}{\mbox{ erg\,s$^{-1}$}}

\newcommand{\gcm}{\mbox{ g\,cm$^{-3}$}}
\newcommand{\etal}{{et al. }}
\newcommand{\kpc}{\mbox{ kpc}}
\newcommand{\mpc}{\mbox{ Mpc}}
\newcommand{\hunits}{\mbox{ km s$^{-1}$ Mpc$^{-1}$}}
\newcommand{\yr}{\mbox{ yr}}
\newcommand{\Lsun}{\mbox{ L$_{\odot}$}}

\newcommand{\gpcden}{\mbox{ Gpc$^{-3}$}}

\newcommand{\gauss}{\mbox{ G}}
\newcommand{\kel}{\mbox{ K}}

\newcommand{\microgauss}{\,\mu\mbox{G}}
\newcommand{\radmsq}{\mbox{ rad m$^{-2}$}}
\newcommand{\kms}{\mbox{ km s$^{-1}$}}
\newcommand{\cmsq}{\mbox{ cm$^{-2}$}}
\newcommand{\pc}{\mbox{ pc}}

\begin{document}
\vskip 0.6 in
\noindent
 
\title{Intergalactic Magnetic Fields from Quasar Outflows}

\author{
Steven R. Furlanetto \& Abraham Loeb
}
\affil{Harvard-Smithsonian Center for Astrophysics, 60 Garden St.,
Cambridge, MA 02138;\\sfurlanetto@cfa.harvard.edu, aloeb@cfa.harvard.edu}

\begin{abstract}

Outflows from quasars inevitably pollute the intergalactic medium (IGM)
with magnetic fields. The short-lived activity of a quasar leaves behind an
expanding magnetized bubble in the IGM. We model the expansion of the
remnant quasar bubbles and calculate their distribution as a function of
size and magnetic field strength at different redshifts.  We generically
find that by a redshift $z\sim 3$, about 5--$20\%$ of the IGM volume is
filled by magnetic fields with an energy density $\gtrsim 10\%$ of the mean
thermal energy density of a photo-ionized IGM (at $\sim 10^4$ K).  As massive
galaxies and X-ray clusters condense out of the magnetized IGM, the
adiabatic compression of the magnetic field could result in the field
strength observed in these systems without a need for further dynamo
amplification.  The intergalactic magnetic field could also provide a
nonthermal contribution to the pressure of the photo-ionized gas that
may account for the claimed discrepancy between the simulated and
observed Doppler width distributions of the Ly$\alpha$ forest. 

\end{abstract}


\section{Introduction}

The interstellar medium of galaxies is known to possess strong magnetic
fields which are dynamically important and reach near equipartition with
the total (turbulent$+$thermal) kinetic energy (Zel'dovich, Ruzmaikin, \&
Sokoloff 1983; Beck \etal 1996).  Substantial magnetic fields are also
known to exist on larger cosmological scales but their dynamical
significance is uncertain. Faraday rotation measurements imply a lower
limit of $1$--$100 \microgauss$ on the tangled magnetic field in the cores
of X-ray clusters (Kim et al. 1990, 1991; Taylor \& Perley 1993; Ge \& Owen
1993; Dreher et al. 1987; Perley \& Taylor 1991), while the comparison
between the synchrotron and inverse-Compton emission by relativistic
electrons in the outer ($\ga 1$ Mpc) envelopes of X-ray clusters indicates
a lower limit of $\sim 0.1\microgauss$ there (Fusco-Femiano et al.  1999;
Rephaeli et al. 1999; Kaastra et al. 1999; Kim et al. 1989).  The observed
field amplitude of $B\ga 10^{-7}$ G in collapsed environments with an
overdensity factor of $\delta \sim 10^3$ translates to $B\ga 10^{-9}$ G at
the mean density of the intergalactic medium (IGM), under the assumption
that the field was adiabatically compressed ($B\propto \delta^{2/3}$) as
the gas collapsed. The corresponding intergalactic magnetic field should
have an energy density of $B^2/8\pi\ga 4\times 10^{-20}~{\rm erg~cm^{-3}}$,
which amounts to $\ga 5\%$ of the thermal energy density of the IGM at the
mean cosmic density and the photo-ionization temperature of $T\sim 10^4$ K,
typical of uncollapsed regions.  The existence of this non-thermal pressure
component could have had a significant effect on the Ly$\alpha$ forest and
on the fragmentation of the IGM into dwarf galaxies.

The origin of the observed magnetic field in galaxies and X-ray clusters is
still unresolved. The dynamo amplification process in galactic disks
requires many dynamical times and cannot account for the tentative
detection of magnetic fields in high-redshift galaxies (Oren \& Wolfe
1995). On the other hand, a cosmic origin in the early universe is
problematic because the comoving scale of causally-connected regions is
small in the pre-recombination epoch (e.g. Quashnock, Loeb, \& Spergel
1989), and field amplification processes are typically weak subsequently
(Harrison 1973; Gnedin, Ferrara, \& Zweibel 2000).  Kulsrud \etal (1997)
proposed a scenario in which the fields are generated by the Biermann
battery mechanism during structure formation and subsequently amplified
through a turbulent energy cascade.  However, as those authors acknowledge,
it is not clear that the field will obtain the observed coherence on
galactic scales after the cascade.

In this paper we explore the possibility that the observed galactic and
intergalactic fields originate from energetic quasar outflows, an idea
originally proposed by Rees \& Setti (1968) for intergalactic fields
and by Hoyle (1969) for galactic fields. These
outflows can carry magnetic flux from the very compact accretion disk ($\la
10^{15}$~cm) around a quasar black hole, where the growth time is very
short, out to cosmological scales ($\sim 10^{24}$ cm), starting at high
redshifts. The stretching of field lines by the outflows could also account
for the coherent field structure which is observed in galactic disks (Daly
\& Loeb 1990; Howard \& Kulsrud 1997; note that supernova-driven outflows
cannot reach the same level of coherence), and explain the large magnetic
flux which is observed in X-ray clusters (\cite{colgate-li}).
Quasar outflows are detected in two forms: (i) radio jets with strong
magnetic fields in about a tenth of all quasars (Begelman, Blandford \&
Rees 1984); 
(ii) broad absorption line (BAL) outflows in a different tenth of the
observed quasar population (possibly due to the covering
fraction of the outflow; Weymann 1997) that are thought to carry magnetic
fields as well (de Kool \& Begelman 1995).  Both types of outflows may
transport a substantial fraction of the accretion energy of the quasar
into the surrounding medium
(see, e.g. Begelman, Blandford, \& Rees 1984; Murray et al. 1995;
Blandford \& Begelman 1999).

It is now known from observations of high-redshift quasars that the diffuse
IGM was reionized at a redshift 
$z \gtrsim 6$ (\cite{fan}).  Most scenarios for
reionization assume that the energy source is radiation from quasars
(Haiman \& Loeb 1998; Valageas \& Silk 1999) or stars (Haiman \& Loeb 1997;
Chiu \& Ostriker 2000; Gnedin 2000), although some authors have also
considered the effects of winds from starburst galaxies (\cite{tegmark}).
Mechanical outflows from quasars could potentially also play a role in this
process.
It is important to examine this possibility, because reionization due to
mechanical outflows (either starbursts or quasars) would have been
qualitatively different from photo-ionization.  Simulations have shown that
reionization in the latter case would have had the character of a phase
transition (Gnedin 2000; Barkana \& Loeb 2001).  This would not be true for
reionization via collisional processes; in such a scenario the flux of
ionizing particles at a given distance from the source is
limited by the particle velocity rather than by the optical depth of
the ambient medium.  Thus, the ionizing agent cannot
efficiently escape pre-ionized regions, and we
would expect reionization to occur over a much broader redshift interval
than in the photo-ionization case.

Because of the short duration of the quasar activity, the ouflow impulse
leaves behind a magnetized high-pressure bubble which expands into the
IGM.  In \S 2 we model the expansion of such a bubble into the surrounding
IGM and add up the cumulative effects of many quasar sources.
In \S 3 we present numerical results from this model, and in \S 4 we
examine their implications for the intergalactic magnetic field.
Throughout the paper we assume a flat universe with a cosmological constant
having $\Omega_0 = 0.3$, $\Omega_{\Lambda0} = 0.7$, $\Omega_b h^2 = 0.019$,
$H_0 = 70 \hunits$, and $\sigma_8 = 0.9$.

\section{Method}

\subsection{Bubble Expansion}

The expansion of an outflow from a quasar into the IGM depends on the
kinetic energy output, the density distribution of ambient gas, and the
gravitational potential well in which the quasar resides.  In our models,
we assume that a fixed fraction of all quasars expel magnetized winds
that carry a fraction of the total quasar radiative luminosity. We
embed each quasar in a 
galactic halo, adopting a simple prescription for the host properties that
has been shown to be succesful in describing the observed quasar luminosity
function at $2\la z\la 5$ (see the review by Barkana \& Loeb 2001, and
references therein).

In the following subsections we describe our model for the expansion of the
magnetized outflow around a quasar.  We first identify the properties of a
quasar host in \S 2.1.1.  We then consider the early expansion
phase, during which the source is still active, in \S 2.1.2.  Once
this active phase ends, the magnetized region relaxes into an
approximately spherical shell.  The overpressured interior forces the shell
to expand, as described in \S 2.1.3.  Finally, in \S 2.1.4, we show how the
expansion depends on the input parameters and assumptions.

\subsubsection{Halo and Environment}

By now, it is widely accepted that quasars are active stages in the growth
of the supermassive black holes at the centers of galaxies.  In modeling the
quasar host properties we adopt the simplest model of Haiman \& Loeb
(1998), whereby each black hole amounts to a fixed fraction of the halo
mass, $\epsilon_h$, and has a universal active lifetime, $\tau_q$,
during which it 
shines at the Eddington limit with a universal spectrum. The two free
parameters of this model can be chosen so as to reproduce the observed
quasar luminosity function at moderate redshifts.  The black hole mass may
be written as
\bq M_{BH} =
\frac{1}{\epsilon_{bol}}\left(\frac{L_B}{1.4 \times 10^{38} \ergs} \right)
\msun,
\label{eq:bhmass}
\eq where $L_B$ is the rest-frame $B$-band luminosity of the quasar,
$M_{BH}$ is the 
mass of the central black hole, and $\epsilon_{bol}$ is the fraction of the
Eddington luminosity radiated in the $B$-band.  The median quasar spectrum
of Elvis \etal (1994) gives $\epsilon_{bol}=0.093$, assuming that the
quasar radiates at the Eddington limit.

Under the assumption that the central black hole of a galaxy contains a
fixed fraction of its total baryonic mass (\cite{magorrian};
\cite{gebhardt}), we express the black hole mass as a fixed fraction of the
total galaxy mass $M_{BH} = \epsilon_h M_{h}$, where $\epsilon_h \approx
(M_{BH}/M_{bulge}) \times (\Omega_b/\Omega_0) \approx 4 \times 10^{-4}$,
assuming the most recent value of $M_{BH}/M_{bulge} \approx 2$--$3 \times
10^{-3}$ (\cite{gebhardt}).

We assume that each quasar begins its active phase at the
formation redshift of its host halo, $z_0$.  
Numerical simulations have shown that relaxed cold dark matter halos have a
universal density profile within their virial radii over a broad range
of masses (\cite{nfw}; 
hereafter NFW).  The dark matter density
profile of the host, $\rho_d(r)$, is 
\bq \rho_{d}(r) = \rho_{c0} (1+z_0)^3
\frac{\Omega_0}{\Omega(z_0)} \frac{\delta_c}{(c r/R_{v}) (1+c r/R_v)^2},
\label{eq:nfwprofile}
\eq
where $\rho_{c0}$ is the critical density today, $\Omega(z_0)$ is
the matter density parameter at redshift $z_0$, and $\delta_c$ is the
characteristic density of the halo, determined by its mass and formation
redshift (NFW).  The concentration
parameter $c$ is a function only of the chosen cosmology and of $\delta_c$, 
\bq
\delta_c = \frac{\Delta_c}{3} \frac{c^3}{\ln(1+c) - c/(1+c)}.
\label{eq:concparam}
\eq
Here $\Delta_c$ is the cosmology-dependent characteristic virial
overdensity (\cite{bryan-norman}).  $R_v$ is the virial radius of the
halo (Barkana \& Loeb 2001): 
\bq
R_{v} = 3.51 \left(\frac{M_{h}}{10^{10} h^{-1} \msun}\right)^{1/3}
\left(\frac{\Omega_0}{\Omega(z_0)} \frac{\Delta_c}{200}\right)^{-1/3}
\left(\frac{1+z_0}{10}\right)^{-1} h^{-1} \mbox{ kpc}.
\label{eq:rvir}
\eq

The collisionless dark matter interacts only gravitationally with the
quasar outflow.  The baryons, on the other hand, interact directly with the
outflow, because the outflow must either sweep the gas aside (during a jet
phase) or sweep it into a shell (if the outflow is spherical).  X-ray
observations 
exclude the existence of extended X-ray halos or cooling flows in most
spiral galaxies in the local universe (Fabbiano 1989).
Hence, we assume that all the gas within the virial radius of the halo
cools onto the central core (disk or bulge) of the galaxy (see \S 2.1.4 for
a discussion of the importance of this assumption).  

Gas infall outside the halo results in a velocity field $v_p(r)$
directed toward the halo.  There is no simple form for this field at low
redshifts in an $\Omega_{\Lambda} \neq 0$ universe, so we make a crude
approximation 
to the self-similar solution for a flat $\Omega_0=1$ universe given by
Bertschinger (1985): 
\bq v_p(r) = \left\{
\begin{array}[l]{lll}
0 & \, & r < R_{v}, \\
\frac{\sigma}{3} \left(\frac{r}{R_{v}} - 4\right) 
& \, & R_{v} < r < 4 R_{v}, \\
\frac{3 H}{2}  \left(r - 4 R_{v}\right) & \, &  4 R_{v} < r < 12 R_{v}, \\
H r & \, & 12 R_{v} < r. \\
\end{array}
\right.
\label{eq:vpec}
\eq 
Here $\sigma$ is the virial velocity dispersion of the halo
(Barkana \& Loeb 2001):
\bq
\sigma = 223
\left(\frac{R_{v}}{h^{-1} \mbox{ kpc}}\right)
\left(\frac{\Omega_0}{\Omega(z_0)} \frac{\Delta_c}{200}\right)^{1/2}
\left(\frac{1+z_0}{10}\right)^{3/2} \kms .
\label{eq:veldisp}
\eq
Inside the accretion shock at the virial radius, the gas is assumed to
be either cold or in hydrostatic equilibrium.  
(In our standard scenario, all of the
gas inside the virial radius has cooled onto the galaxy, so the second
possibility comes into play only when we examine the
dependence of the model on our assumption of cooling in \S 2.1.4.)
Just outside of this radius gas falls onto the halo, while farther away
the gas recedes with the Hubble flow.  The 
precise form of the infall velocity field has only a small
effect on the final results ($\lesssim 5\%$).  The self-similar
solution also determines the density distribution of the dark matter and
gas through which the outflow must travel (Bertschinger 1985).
Outside of the accretion 
shock both the dark matter and gaseous components follow
the profile $\rho
\propto r^{-2.25}$ for $\Omega_b\ll \Omega_0$.  There are two important 
simplifications inherent to our treatment of this infall region.  
First, the self-similar
solution applies to an $\Omega_0=1$ universe, while we use 
an $\Omega_{\Lambda} \neq 0$ cosmology.  However, because quasars are
most common at 
high redshifts where the effects of the cosmological constant are small,
this introduces only small changes in our results.
Second, we ignore the slow growth of
the infall region as the halo accretes more mass.  This is justified
because those outflows that escape the gravitational potential of their
host traverse the infall region quickly, before that region has evolved
significantly.  

We find in our calculations that the detailed density structure of the halo
infall region has little effect on the results.  Although
the density solutions of Bertschinger (1985) and  NFW differ
substantially within the virial radius, the accumulated mass in the
expanding shell is dominated by swept-up mass at large radii, at which the
density is close to the cosmological mean in both cases.  For computational
convenience, we assume that the dark matter continues to follow the NFW
profile outside of $R_v$ until $\rho_d$ falls below the mean cosmological
density at $z_0$.  Outside of this radius, we assume that the density field
follows the (time-dependent) mean cosmological density.  The simple
prescription that the gas simply follows the dark matter, with 
a density $\rho_g =
(\Omega_b/\Omega_0) \rho_d$, results in only a minor loss of accuracy
compared to the calculation with the full 
density profile of Bertschinger (1985).

We have now associated each quasar of rest-frame $B$-band luminosity $L_B$
with a host halo.  Next, we assign a kinetic (or mechanical) luminosity
$L_K = \epsilon_K L_B$ to each quasar.  Recent evidence suggests that the
kinetic and radiation luminosities are comparable for radio jets. Willott
\etal (1999) have examined the correlation between narrow-emission line
luminosity (which is excited by radiation from the
central engine, and thus is related to the bolometric luminosity
$L_{bol}$) and radio luminosity (assumed 
to correlate with the jet power) in radio galaxies, over three decades in
radio power.  They found that their theoretical model, in which $L_K
\propto L_{bol}$ and $0.05 \lesssim L_K/L_{bol} \lesssim 1.0$, was
consistent with the available data.  Because $L_{bol} \approx 10 L_B$ in
the rest-frame median quasar spectrum of Elvis \etal (1994), we 
conservatively adopt $\epsilon_K = 1.0$ for radio-loud quasars.

In broad absorption line (BAL) quasars, the relation between radiative and
kinetic power is even more uncertain.  Given the column density of
absorbing gas $N_H$, the mean outflow velocity $v_{BAL}$, the radius of the
absorbing system $R_{BAL}$, and its covering fraction of the central source
$f_c$, the kinetic luminosity of the system is \bq L_K \approx 2 \pi f_c
N_H m_p v_{BAL}^3 R_{BAL}, \eq where $m_p$ is the proton mass.  Of these
quantities, the best known is $v_{BAL} \lesssim 0.1 c$, which can be
directly inferred from the absorption lines.  Unified models suggest that
$f_c \sim 0.1$, although this value is highly uncertain and could vary
among individual sources (\cite{weymann}).  The radius $R_{BAL}$ can be
constrained based on the photo-ionizing flux of the quasar, and estimates
for individual sources cover the range $R_{BAL} \sim 1 - 500 \pc$
(\cite{krolik}; \cite{turnshek}). The value of $N_H$ can be constrained
with X-ray absorption data. ASCA measurements have recently indicated much
higher column densities of absorbing gas than previously suspected, $N_H
\approx 10^{22}-10^{23} \cmsq$ (\cite{gallagher}).  Given that the inferred
column density of this gas is correlated with that of the UV-absorbing gas
(Gallagher et al. 2000), our estimates for individual quasars range from
$\epsilon_K \sim 0.01 - 10$.  Again, $\epsilon_K = 1.0$ appears to be a
reasonable choice.

\subsubsection{Early Expansion and Isotropization}

Expansion during the initial phase, while the quasar is active, depends
strongly on the geometry of the outflow.  For a radio-loud quasar (RLQ),
the outflow is tightly collimated, as observed in nearby galaxies.  The
geometry of a BAL quasar outflow is unknown.  For concreteness, we begin
with a discussion on outflow in jets.

During the active lifetime of a radio loud quasar, we assume that the
outflow is collimated in twin jets.  The material in radio jets is known to
be highly relativistic near the central source, and at least moderately
relativistic away from the nucleus (\cite{bbr}; \cite{tingay};
\cite{biretta}).  The momentum flux in each jet is therefore $L_K/(2 c)$.
As the jet strikes the ambient medium, it forms a contact discontinuity
between forward and reverse shocks (\cite{bbr}). The expansion rate of the
contact discontinuity can be found by balancing the outward momentum
flux of the
particles in the jet with the external ram pressure of the medium: \bq
\frac{L_K}{2 A_s c} = \rho_g (v_s-v_p)^2,
\label{eq:rampressure}
\eq where $A_s$ is the total area of the interface (generally larger than
the instantaneous area of the jet) and $v_s$ is the velocity of the
discontinuity.  We assume that the shock front has a constant
half-opening angle $\theta = 5^{\circ}$ relative to the central source
(\cite{bridle-perley}); this could correspond either to a shock front with
a constant transverse expansion rate or a precessing, narrow jet.  Note
that observations show that $v_s \ll c$ under the conditions of interest
here (\cite{krolik}).  This may also be written as
\begin{equation}
\frac{dR}{dt} - v_{p} = 5.5 \times 10^{-3}
\left(\frac{5^{\circ}}{\theta}\right) \left(\frac{L_K}{10^{46}
\ergs}\right)^{1/2} 
\left(\frac{\rho_g(R)}{10^{-29} \gcm}\right)^{-1/2}
\left(\frac{R}{\mbox{kpc}}\right)^{-1} \mbox{
}\frac{\mbox{kpc}}{\mbox{yr}},
\label{eq:jetprop}
\end{equation}
where $R$ is the length of the jet (measured from the center of the host
galaxy).  

Once the quasar becomes dormant, we assume that the jets rapidly balloon
outward to form a sphere with a radius equal to the final length of the
jet.  We then assume that the total mechanical outflow energy, $E_K = L_K
\tau_q$, goes into the thermal energy of the bubble.  Pressure gradients
within this region will rapidly force most of the gas into a shell at the
edge.  The energy that remains inside the shell produces a hot,
overpressured interior.  During the formation of the thin shell, some of
the thermal energy of the interior is lost to ionization processes
and radiation.  In
the case of supernova remnants, $\sim 25\%$ of the explosion energy remains
in the thermal energy of the interior after the transition to the shell
phase, while another $\sim 25\%$ is contained in the kinetic energy of the
shell (\cite{cox}; \cite{falle}).  In our scenario, the remnants are
considerably more rarefied, and so a much smaller fraction of the energy
would be radiated away during the formation of the shell.  [Note that in a
cosmological context, even adiabatic blastwaves quickly form thin shells
(\cite{ost-mckee}).]  The uncertainty about the energy loss during this
process can be absorbed into the much larger uncertainty in $\epsilon_K$
(the significance of which will be discussed in \S 2.1.4).

The assumption of spherically symmetric expansion may not be valid for all
quasar environments.  Because halos tend to form at the intersection of
sheets and filaments, the density distribution around the halos will be
inhomogeneous.  The shells will therefore expand along the path of least
resistance and preferentially in directions away from these filaments
(\cite{martel}).  This is in fact an important effect; taken literally, our
model would suppress infall in all but the largest halos as the shell
sweeps away the infalling matter.  In reality, however, infall
primarily occurs in channels along filaments and sheets.  In these
directions, the expansion of the shell will most likely be halted quickly
and infall will continue.  Meanwhile, in directions away from the
filaments, the shell will continue to expand.  Although this process will
certainly affect the geometry of the resulting magnetic field, it may not
strongly affect the overall filling factor given by our model.  Including
such geometric effects will, in any case, require the use of numerical
simulations.

We also assume that the quasar expels magnetic energy during its active
phase.  Unfortunately, the behavior of the magnetic fields within jets is
not well-understood.  Flux conservation would cause the fields to stretch
and weaken as they propagate outward, but at least in some powerful jets
the fields are amplified during the jet propagation (\cite{bbr}).  We
ignore the subtleties involved in the jet evolution and simply assume that,
at the end of the the jet phase, the bubble has magnetic energy equal to a
fixed fraction of the injected thermal energy, so that $E_B = \epsilon_B
E_K$.  The value of $\epsilon_B$ is highly uncertain, and we will regard it
as a free parameter (see \S 3).  The geometry of the field is also unclear;
in radio-weak jets the transverse field dominates far from the core, but
the opposite is true in radio-strong jets (Bridle \& Perley 1984), because
of the aforementioned field amplification processes within such jets.  We
therefore assume for simplicity that the field is tangled on small scales
and that it affects the shell dynamically only through an isotropic
pressure component $p_B = (1/3) B^2/8 \pi$.

Our prescription implicitly assumes that the magnetized plasma ejected from
the quasar mixes efficiently with the ambient thermal gas.  Such mixing
has been inferred in some systems.
For example, Bicknell, Cameron, \& Gingold (1990) showed that the
fluctuating Faraday rotation signal observed in Cygnus A and PKS 2104-25N
may be due to a region in which the magnetic field of the radio source has
mixed with entrained ambient gas through turbulence excited by the
Kelvin-Helmholtz instability on the surface of the radio lobes.

In BAL quasars, the geometry of the outflow and its covering fraction are
unknown.  In \S 2.1.4, we will show that the extreme assumption of a
spherically symmetric outflow during the early phase results in only a
small decrease in the final radius relative to the jet case.  Because
models suggest that $f_c \sim 0.1$ in BAL quasars (\cite{weymann}), so that
the outflow is closer to being collimated than to being isotropic, we use
the jet propagation algorithm in our calculations for BAL quasars as well
as for RLQs with only small errors.

\subsubsection{Late Expansion}

The isotropization of the energy burst from the quasar results in an
overpressured shell embedded within a smooth, lower pressure environment.
The shell expands, sweeping up ambient matter, until its velocity matches
the Hubble flow (\cite{tegmark}) or until the binding gravitational
potential of the halo causes the shell to recollapse.  The equation of
motion of the shell, with radius $R$ and mass $M_s$, is
\begin{eqnarray}
\ddot{R} = \frac{4 \pi R^2}{M_{s}} (p_T + p_B - p_{ext}) - \frac{G}{R^2}
\left(M_{d}(R) + M_{gal} + \frac{M_{s}}{2} \right) \nonumber \\ +
\Omega_{\Lambda}(z) H^2(z) R - \frac{\dot{M}_s}{M_{s}} [\dot{R} -
v_p(R)],
\label{eq:motion}
\end{eqnarray}
where $H(z) = H_0 h(z) = H_0 \sqrt{\Omega_0(1+z)^3 + \Omega_{\Lambda 0}}$,
$\Omega_{\Lambda}(z) = \Omega_{\Lambda 0} [\Omega_0(1+z)^3 +
\Omega_{\Lambda 0}]^{-1}$, $p_T$ is the thermal pressure of the bubble
interior, $p_{ext}$ is the pressure of the external medium, $M_{d}(R)$ is
the mass of dark matter interior to the shell, $M_{gal}$ is the mass of
baryonic matter that has cooled onto the galaxy, and $\dot{M}_s$ is
the rate at which mass is swept into the shell (see below).  
Here the first term describes the pressure
gradient across the shell, the second term describes the gravitational
attraction between 
the halo and the shell (and the self-gravity of the shell), and the third
term describes the effect of the cosmological constant, which amounts to a
fraction $\Omega_{\Lambda 0}$ of the critical density at present.  The
final term describes the drag force from accelerating the swept-up matter
to the bubble velocity.

As the shell travels outward it sweeps up ambient matter at a rate
\bq
\dot{M}_s = \left\{
\begin{array}[l]{ll}
0 & v_p(R) \geq \dot{R}, \\
4 \pi R^2 \rho_g(R) [\dot{R} - v_p(R)] & v_p(R) < \dot{R}. \\
\end{array}
\right.
\label{eq:mshelldot}
\eq

Flux conservation of the tangled magnetic field during the expansion of
the shell implies 
\bq p_B(R) = p_B(R_i) \left(\frac{R_i}{R}\right)^4,
\label{eq:pbevol}
\eq where $R_i$ is the initial radius of the shell.  We implicitly assume
here that the magnetic field is frozen into the expanding bubble, i.e. that
the bubble is ionized.  In fact, the plasma in individual bubbles may
recombine, after which ambipolar diffusion could separate the neutral gas
from the magnetic field.  However, we find that recombination in the
bubbles does not occur until cosmological
reionization (see \S 3.3), and so we can safely ignore this possibility.

The thermal pressure decreases both due to expansion and due to energy
losses (\cite{tegmark}): \bq \dot{p}_T = \frac{\Lambda}{2 \pi R^3} - 5 p_T
\frac{\dot{R}}{R}.
\label{eq:ptevol}
\eq Here $\Lambda$ is the complete heating/cooling function, which
includes several components due to different physical processes
inside the bubble interior, \bq
\Lambda = L_{diss} - L_{brem} - L_{comp} - L_{ion}.
\label{eq:lumsum}
\eq The second and third terms account for radiative cooling: $L_{brem}$
describes cooling due to bremsstrahlung emission and $L_{comp}$ describes
inverse-Compton cooling off the cosmic microwave background.  

In our model, the ambient medium is swept up and
accelerated to the shell expansion speed through inelastic
collisions with the shell material.  $L_{diss}$ describes the fate of
the energy dissipated in this 
process.  The energy could either be radiated away within the
shell (e.g., during shock cooling), or injected into the bubble interior
through turbulence.  We let $f_d$ be the fraction of the total kinetic
power that is injected as heat into the interior, 
\bq L_{diss} =
\frac{1}{2} f_d \dot{M}_{s} [\dot{R} - v_p(R)]^2.
\label{eq:fdiss}
\eq For a fully radiative blastwave, $f_d=0$ (\cite{ost-mckee}).  For an
energy-conserving (adiabatic) blastwave, the amount of energy injected into
the interior can be estimated in the following manner.  Consider a uniform
shell (of thickness $\ell \ll R$) sweeping up matter at a shock as it
travels through an ambient medium of density $\rho_0$.  Let the material
inside the shell have a pressure $P_1$ and a mass density $\rho_1$.  Inside
the cavity created by the shell, there is a rarefied, hot bubble of
approximately uniform pressure $P_2$.  If we assume that the shell retains
most of the swept-up matter, then the shock jump conditions
($\rho_1=4\rho_0$) and mass conservation imply that for a stationary ambient
medium, $\ell \approx R/12$.  In the cosmological case, the shell is even
thinner (Ostriker \& McKee 1988; Ikeuchi \etal 1983). At the inner boundary
of the shell, $P_1 \approx P_2$. Hence, the ratio between the thermal
energy of the bubble cavity, $E_2$, and the thermal energy of the
shell itself, $E_1$, is 
\bq \frac{E_2}{E_1} = \frac{P_2 V_2}{P_1 V_1} \gtrsim 4,
\label{eq:fdcal}
\eq where $V_1$ and $V_2$ are the volumes of the shell and cavity,
respectively.  Thus, $E_2/(E_1+E_2) \gtrsim 80\%$ of the blastwave energy
resides in the cavity, and $f_d \gtrsim 0.8$ for an adiabatic blastwave.

In order to identify whether the shells are adiabatic or radiative, we
consider the effect of cooling just behind the shock front.  The shock is
initially strong, so that the postshock density is $\rho_s = 4
\rho_g(R)$ and the postshock temperature is $T_s = (3 \mu m_p
v_s^2)/(16 k)$, where $v_s=\dot{R} - v_p$ and $k$ is Boltzmann's
constant.  The cooling time for this region is \bq \tau_c =
\frac{3 \mu_b m_p k T_s}{\rho_s \Lambda_s(T_s)} ,
\label{eq:cooltime}
\eq assuming that the gas is fully ionized.  Here $\mu_b m_p$ is the mean
mass of the ions and $\Lambda_s$ is the cooling function of the shocked
material within the shell.  We use the zero-metallicity cooling function
compiled by Sutherland \& Dopita (1993), with the addition of Compton
cooling.  When the age of the bubble exceeds $\tau_c$, we assume that the
bubble makes an instantaneous transition from an adiabatic ($f_d = 0.8$)
to a radiative ($f_d=0$) phase.  Our calculations show that the
intergalactic bubbles begin as adiabatic blastwaves and remain so over a
substantial fraction of their subsequent expansion (c.f. \cite{voit}).  

We find that a fixed value of $f_d=0.8$ for the entire bubble history is an
adequate approximation, and adopt this value throughout.  For very
high-redshift quasars ($z_0 \gtrsim 15$), this prescription overestimates
the final bubble size by $\sim 10\%$; however, the error decreases with
decreasing $z_0$ and is negligible for $z_0 \lesssim 5$.  This reflects the
fact that Compton cooling dominates within the shell, so that $\Lambda_s
\propto (1+z_0)^4$.  Quasars at very high redshifts make only small
contributions to our final results so the errors introduced by fixing
$f_d=0.8$ are small.  Because $f_d$ has a relatively strong effect on the
final radius (see \S 2.1.4), and because the blast wave could in principle
have a shorter cooling time due to clumpiness, we also show some results
for $f_d=0$.

The strong shock surrounding the expanding bubble could also accelerate
cosmic rays and relativistic electrons (\cite{eichler}), similarly to
supernova remnants (\cite{koyama}; \cite{tanimori}).  These accelerated
particles would remain in the bubble interior and provide an extra
non-thermal pressure force on the shell, although this pressure would be
reduced if the relativistic cosmic rays leak out from the interior.  If
leakage is small, such a pressure component could mimic a non-zero value of
$f_d$ even at late times.

Finally, some or all of the swept-up material will be ionized.  Stebbins \&
Silk (1986) found that if the temperature $T > 1.5 \times 10^4 \kel$,
hydrogen will be collisionally ionized; below this critical value the gas
remains neutral.  If $f_d=0.8$, the shock will ionize all of the
incoming hydrogen provided that the postshock temperature exceeds this
threshold.  Even if this condition is not satisfied, the bubble interior may be
hot, and so we follow the temperature evolution of the initially ionized
bubble, \bq T_b = \frac{p_T}{n_b k},
\label{eq:bubbletemp}
\eq
where $n_b$, the particle number density inside the bubble, is
\bq
n_b = \frac{3 \epsilon_i f_m}{4 \pi \mu_b m_p} \frac{M_{s}}{R^3}.
\label{eq:bubbledensity}
\eq 
Here $\epsilon_i= 1
\,(2)$ if the gas is neutral (ionized).  We assume
that a fraction $f_m\sim 0.1$ of the shocked shell material 
leaks into its interior; this is likely to be an overestimate,
particularly at late times (\cite{ost-mckee}; \cite{ikeuchi}),
but it has little effect on the results.  The energy lost 
per unit time in ionizing the incoming hydrogen is then approximately 
\bq
L_{ion} = E_H \times \frac{f_i \dot{M}_s}{\mu_b m_p}
\label{eq:ionloss}
\eq
where $E_H = 13.6 \mbox{ eV}$ is the ionization potential of
hydrogen and 
\bq
f_i = \left\{
\begin{array}[l]{ll}
1 & f_d=0.8,\,z>z_r, T_s > 1.5 \times 10^4 \kel \\
0.1 & z>z_r,\, T_s < 1.5 \times 10^4 \kel,\, T_b > 1.5 \times 10^4 \kel \\
0 & \mbox{otherwise}
\end{array}
\right.
\label{eq:fidefn}
\eq
is the fraction of swept-up material that is ionized.
Note that if the universe has already been
reionized at $z=z_r$, no ionization losses occur.  In principle, once
the bubble has 
expanded sufficiently, the temperature could decrease to a point at which
hydrogen recombines.  When this occurs, there will be a rapid pressure drop
inside the bubble.  However, we can neglect this effect because for nearly
all bubbles in our model, recombination  does not occur until well after the IGM has been reionized.

Our model requires knowledge of the external pressure as a function of
redshift.  Prior to reionization, the IGM around the quasar may recombine
shortly after the quasar stops illuminating it with ionizing photons. In
this case the bubble would propagate into a neutral IGM where the pressure
is very low.  The baryon temperature decreases adiabatically,
$T_{IGM} \propto (1+z)^2$ between the redshift of thermal decoupling
from the microwave background ($z_t\approx 100$) and reionization
(Barkan \& Loeb 2001).  We
assume that reionization occurs instantaneously at $z_r=7$ (Gnedin 2000),
and that it raises the IGM temperature to $T_{IGM} \approx 1$--$2 \times
10^4 \kel$  subsequently. 
The IGM temperature continues to rise slowly until the present time,
particularly in moderately overdense regions (\cite{dave};
\cite{gnedin}). In modeling $z<z_r$, we adopt $p_{ext} = 2 n(z) k T_{IGM}$, 
where $T_{IGM} = 1 \times 10^4 \kel$ and $n(z)$ is the mean 
cosmic baryon density.
  The
effects of a higher ambient pressure are examined in \S 2.1.4.
Prior to reionization, the quasar photo-ionizes a large region of the IGM
in its vicinity, and the shell will thus initially propagate through an
ionized medium.  In this case, the external pressure will not be
negligible even at high redshifts; however, the source will also not
lose any internal energy in 
ionizing the swept-up material.  We find, like Barkana \& Loeb (2001), that
these two effects nearly cancel, and so we ignore this subtlety.

Equations (\ref{eq:motion}), (\ref{eq:mshelldot}), (\ref{eq:pbevol}),
and (\ref{eq:ptevol}) fully describe the growth of each bubble.  

For massive halos, the deep gravitational potential of the host galaxy
prevents the shell from escaping and causes it to fall back into the
virialized region.  For $\epsilon_B = 0.1$, $\tau_q = 10^7 \yr$, and
$f_d = 0.8$, we find that the maximum halo mass for which escape is
possible is $M_{h,max} \approx 2 \times 10^{13} \msun$ at $z = 3$ and
$M_{h,max} \approx 3 \times 10^{11} \msun$ at $z = 15$.  
We do not include halos more massive than this in our calculation;
their magnetized bubbles do not experience any cosmological expansion and
therefore make only a small contribution to the volume filling factor of
magnetized regions.  The shells in less massive halos continue to expand in
physical coordinates until the present day. We halt such shells at the
point of their maximum expansion in comoving coordinates.

\subsubsection{Parameter Dependences}

A simple, and surprisingly accurate, estimate of the final bubble
radius can be obtained by 
noting that the mechanical (plus magnetic) quasar energy $E_0$ will
essentially accelerate all the matter it encounters into a thin shell at a
comoving distance $\hat{R}_{\rm max} = R_{\rm max}(1+z)$ from the central
source.  In doing so, it must accelerate this material to the Hubble flow
velocity at that distance, $v = H(z_0) R_{\rm max}$.  (Here we assume
that the bubble reaches its maximum comoving size quickly compared to the
age of the universe; see below for a discussion of the importance of
this assumption.) 
At the onset of shell expansion, there are two energy reservoirs: the
input energy of the quasar itself, $E_0$, and the kinetic energy of the
pre-existing Hubble flow within a distance $R_{\rm max}$ of the quasar,
$(3/10) M_s v^2$.  If the shell is radiative, the Hubble flow
energy is lost during shock cooling and is unavailable
in expanding the shell.  If the expansion is adiabatic, a fraction
$f_d \sim 0.8$ of the Hubble
flow energy remains in the bubble and contributes to the expansion.
Therefore, if we neglect the gravitational
deceleration of the host halo as well as cooling inside the bubble
cavity, energy 
conservation implies $E_0 \approx K M_s v^2/2$, where $K = 0.52$
if $f_d=0.8$ and $K=1$ if $f_d=0$. 
Equivalently,
\begin{eqnarray}
\hat{R}_{\rm max} \approx \frac{1.3}{K^{1/5}}
\left[ \left(\frac{1+z_0}{h(z_0)}\right)\left( \frac{0.7}{h}
\right)\right]^{2/5} \left[\left(\frac{\epsilon_{bol}}{0.093} \right)\left(
\frac{\epsilon_{h}}{4 \times 10^{-4}}\right) \left(\frac{\epsilon_K (1+
\epsilon_B)}{1.1}\right)\right]^{1/5} \times \nonumber \\
\left[\left(\frac{0.019}{\Omega_b h^2}
\right)\left(\frac{\tau_q}{10^7 \yr}\right)\left(\frac{M_h}{10^{10} \msun}
\right)\right]^{1/5} \mpc .
\label{eq:radestimate}
\end{eqnarray} 
Note that the resulting $\hat{R}_{\rm max}$ is only a weak function of $E_0
\propto \epsilon_K (1+\epsilon_B) \tau_q M_h$, and so is rather robust to
changes in our model parameters.

We can also estimate the maximum halo mass for which the escape of the
bubble shell is 
possible by comparing the total gravitational potential energy of
the gas in an NFW halo (out to $r=\infty$) to the total energy output of
the quasar plus the Hubble flow kinetic energy of the swept-up medium,
if it is not radiated away through shock cooling.  The result is that
the shell can escape if $M \lesssim M_{esc}$, where
\begin{eqnarray}
M_{esc} = \frac{1.4 \times 10^{13}}{K^{3/2}}
\left[
\left(\frac{\epsilon_{bol}}{0.093} \right)\left(
\frac{\epsilon_{h}}{4 \times 10^{-4}}\right) \left(\frac{\epsilon_K (1+
\epsilon_B)}{1.1}\right)\left(\frac{\tau_q}{10^7 \yr}\right)
\right]^{3/2}
\nonumber \\ \times
\left[
\left(\frac{1+z_0}{11}\right)
\left(\frac{\Omega_b h^2}{0.019}\right)
\right]^{-3/2}
\left(\frac{\Omega_0}{\Omega(z_0)}\right)
\left(\frac{\Delta_c}{200}\right)^{-1/2}
g(c)^{-3/2} \msun.
\label{eq:massest}
\end{eqnarray}
In this expression, $g(c)$ is a function of the concentration
parameter of the halo [which must be calculated numerically for each
halo; see equation (\ref{eq:concparam})]:
\bq
g(c) = \frac{c}{[\ln(1+c) - c/(1+c)]^2}.
\label{eq:gc}
\eq
The above expression provides a crude estimate of 
the halo mass above which gravity
dominates the shell dynamics.  At $z=15$, it overestimates the 
actual threshold mass by a factor of $\sim
4$; at $z=5$ it underestimates this mass by a factor of $\sim 2$.

Figure 1 shows the final ($z=0$) bubble radius $\hat{R}$ as a function of
halo mass.  The solid curve shows the results of our numerical model, with
$f_d=0.8$, while the dotted line is calculated from equation
(\ref{eq:radestimate}) with $K=0.52$ for $M < M_{esc}$.  Both curves assume
$z_0=10$, $\tau_q = 10^7 \yr$, and $\epsilon_B = 0.1$.  The agreement is
surprisingly good.  Over most of the mass range, equation
(\ref{eq:radestimate}) underestimates the final comoving radius by only
$\sim 20\%$.  This accuracy is in fact due to a fortuitous cancellation of
two errors in our estimate.  First, we neglect cooling within the bubble
cavity, the external pressure of the IGM, and the gravitational potential
of the host halo.  For these reasons we would expect equation
(\ref{eq:radestimate}) to overestimate the final comoving radius.  On the
other hand, the estimate also neglects cosmological deceleration.  As the
age of the universe increases, the Hubble flow decelerates (until the
cosmological constant begins to dominate at $z \sim 1$).  Therefore, the
Hubble flow velocity at a fixed comoving distance from the host decreases
with decreasing redshift. Because the mass within a given comoving radius
is fixed, \emph{less} energy is required in order to accelerate the shell
to a given comoving distance at \emph{smaller} redshifts.  Equation
(\ref{eq:radestimate}) assumes that the shell reaches its maximum radius
instantaneously, while in reality the shell continues to expand for many
Hubble times (see Figure 3).  For this reason, we would expect it to
underestimate the final radius.  Figure 1 shows that these two errors
nearly cancel.  Voit (1996) constructed an analytic self-similar solution
to the expansion of a blastwave in an $\Omega_{\Lambda}=0$ universe that
included the effects of cosmological deceleration.  Barkana \& Loeb (2001),
using a numerical model similar to ours, show that this solution
overestimates the true size of the blastwaves by a factor of $\sim 2$,
because it neglects cooling, external pressure, and the gravity of the host
halo.  Evidently, ignoring cosmological deceleration gives a factor of 2.5
underestimate, and so our estimate is only $\sim 20\%$ smaller than the
true bubble size.

Figure 2 shows the numerical solution for the comoving bubble radius
$\hat{R}$ as a function of redshift for several different examples.  All
cases assume that $M_h = 10^{10} \msun$, $z_0 = 10$, $\tau_q=10^7 \yr$, and
$\epsilon_B=0.1$.  Results are shown for our standard scenario (solid
line), a case in which the gas in the halo of the host galaxy does not cool
(short dashed line), a case in which $T_{IGM} = 2 \times 10^4 \kel$ after
reionization (long dashed line), and a case in which the outflow is
initially spherical rather than a jet (dot-dashed line); all of these
curves assume $f_d=0.8$.  Also shown is a case in which $f_d=0$ (dotted
line).

Clearly, the parameter which has the greatest effect on the final result is
$f_d$.  Varying our other 
assumptions makes considerably less of a difference in our 
results.  Although the initial expansion slows down considerably when the
virialized gas does not cool, the late-phase
expansion remains roughly the same because the swept-up mass is dominated
by gas outside the virial radius, so that the final radius is only
slightly smaller.  To describe spherical expansion during the
initial phase, we 
use the model of \S 2.1.3, including an
extra source term in $\Lambda$ for the quasar mechanical luminosity.  This
has a somewhat larger effect because plowing through the ambient
medium during this phase is significantly more difficult than sweeping
aside the 
matter in a jet beam.  As a result, spherical flows decelerate more rapidly
early on, decreasing the final bubble radius by $\sim 5 \%$. 

Increasing the post-reionization temperature to the value typical of quasar
reionization models, $T_{IGM} = 2 \times 10^4 \kel$,
decreases the final radius 
by $\sim 5\%$.  In any reionization model, the temperature of the IGM
is expected to continue to rise 
until the present day in moderately overdense regions (Dav\`{e} \etal
1999).  This is particularly important as groups of galaxies form at
$z \lesssim 2$; the gas between the group galaxies will be at the
virial temperature of the group, so outflows from quasars embedded in
these systems must propagate through a hot, high-density medium.  This
pressure increase could suppress 
the filling factor of magnetized regions from low-$z$ quasars, but it
has little effect on the 
high-$z$ contribution because the contributing halos will have already
expanded to be very close to their maximum comoving size by the time the
mean IGM temperature increases substantially.  The effects of  an
increase in the external pressure 
will be discussed further in \S 3.1.

Figure 3 shows the comoving bubble radius $\hat{R}$ as a function of
redshift for quasars which produce an outflow at different redshifts $z_0$.
In all cases $\tau_q=10^7 \yr$, $\epsilon_B=0.1$, and $f_d=0.8$.  The solid
curves assume $M_h = 10^{10} \msun$ with $z_0 = 20,15,10,5, \mbox{ and }
3$, from right to left.  The dashed curve shows the bubble radius for $M_h
= 3 \times 10^{12} \msun$ and $z_0 = 10$, illustrating the evolution of a
quasar outflow in a massive halo in which the gravitational potential
suppresses the expansion into the IGM.  As expected from equation
(\ref{eq:radestimate}), older quasars produce \emph{smaller} bubbles.  This
is simply a result of cosmological deceleration: high-redshift outflows
join the Hubble flow earlier, when the expansion speed for a fixed comoving
distance is larger.  Therefore, accelerating a shell to a fixed comoving
size requires a more energetic outflow (or a more massive halo, according
to our prescription) at higher redshifts.  Note that the $z_0=3$ quasar has
not yet approached its asymptotic final radius.

\subsection{Source Populations}

The most easily detectable type of quasar outflow is a radio
jet. Unfortunately, the relation between the radio power and kinetic jet
power is complicated and highly uncertain (\cite{bbr}; \cite{willott}), and
so it is difficult to infer a kinetic luminosity function from the radio
luminosity function of quasars.  In addition, other forms of outflows, such
as those in BAL quasars, may not be radio loud. On physical grounds, one
would expect the kinetic luminosity to carry some fraction of the total
accretion energy of the quasar (e.g., Blandford \& Begelman 1999), the rest
of which is carried by radiation emanating from the inner region of the
accretion flow.  We therefore calibrate the kinetic luminosity of a quasar
based on its bolometric luminosity, which in turn can be inferred (through
the appropriate $k$-correction for the universal spectrum of Elvis et
al. 1994) from its rest-frame $B$-band luminosity (see \S 2.1.1).

For the optical luminosity function of quasars, we adopt an empirical
parametrization based on observations at low redshifts ($z\la 4$), and
extrapolate it to higher redshifts using a simple theoretical model.  Pei
(1995) presents two fits to the observed luminosity function at low
redshifts: the first is a double power-law, which provides a divergent
contribution as the quasar luminosity approaches zero, and the second is a
modified Schechter function, which converges for small luminosities.
Because small halos have a large effect on the results, we conservatively
use the latter form.  We modify equation (9) of Pei (1995) to our chosen
cosmology in order to describe the quasar abundance as a function of halo
mass and redshift, $\Phi(M_h,z)$.  Since $\tau_q \ll H(z)^{-1}$, the
formation rate per comoving volume of quasars with outflows is \bq
\frac{dn_q}{dM_h dz} \approx f^{Pei} \frac{\Phi(M_h,z)}{\tau_q} \left|
\frac{dt}{dz} \right| \qquad \qquad z \leq 4,
\label{eq:peinumber}
\eq where $f^{Pei}$ is the fraction of quasars with magnetized
outflows.

We extrapolate the quasar luminosity function to high redshifts, $z\ga 4$,
using the Press-Schechter mass function of halos (\cite{press}).  Previous
studies have shown that under the assumptions of a universal quasar
lifetime and a universal quasar light curve, this approach can reproduce
the observed quasar $B$--band luminosity function at $2 \lesssim z
\lesssim 5$ (\cite{haiman-loeb}).  The number of quasars with outflows
per comoving 
volume forming in each host mass range at a redshift $z$ is then \bq
\frac{dn_q}{dM_h dz} \approx f^{PS} \frac{d}{dz} \left( \frac{dn_{PS}}{d
M_h} \right) \qquad \qquad z > 4,
\label{eq:psnumber}
\eq where $dn_{PS}/dM_h$ is the Press-Schechter mass function and $f^{PS}$
is the fraction of these halos hosting magnetized outflows.  

Matching the two formation rates at $z \approx 4$ yields the relation 
\bq
f^{Pei} = f^{PS} \left(\frac{\tau_q}{10^6 \yr}\right).
\label{eq:matchcounts}
\eq
Therefore, we can either adopt a quasar lifetime
of $\tau_q = 10^6 \yr$ and choose $f^{PS} = f^{Pei}$, as shown by Haiman
\& Loeb (1998), or adopt $\tau_q = 10^7 \yr$ and $f^{PS} = 0.1
f^{Pei}$.  We choose the latter possibility, because spectral aging of
radio lobes indicates ages of $\sim 1$--$3 \times 10^7 \yr$ (\cite{kaiser}).

The minimum halo mass to host a quasar, $M_{h,min}$, is determined by the
criteria for efficient cooling and infall.  Before reionization, the lack
of metals in the primordial gas prohibited efficient cooling unless the
virial temperature $T_v > 10^4 \kel$, so as to allow atomic excitations and
cooling via line radiation (Barkana \& Loeb 2001).  After reionization, the
increased pressure due to photo-ionization heating increases the Jeans
mass, suppressing infall onto low-mass galaxies.  This process prevented
the gas from collapsing into halos with a circular velocity $v_c \lesssim
50 \kms$ (Efstathiou 1992; Navarro \& Steinmetz 1997; Thoul \& Weinberg
1996).  We also allow for the possibility of a low-mass cutoff for the
hosts of quasars with strong outflows.  Laor (2000) has argued for such a
relation in radio-loud quasars on empirical grounds, although there is no
firm physical theory for the existence of such a cutoff.

\section{Results}

In most of our calculations we assume $\epsilon_B = 0.1$ and $f_R^{Pei} =
0.1$, the latter being the typical fraction of radio-loud quasars
(\cite{stern}).  We refer to this as the \emph{RLQ model}.  The magnetic
energy fraction in radio jets is highly uncertain; estimates are commonly
based upon an assumption of equipartition, although there is a notorious
lack of observational data to support this assumption (\cite{bbr}).  At
equipartition, the total energy is shared between hydrodynamic motions,
thermal protons, thermal electrons, accelerated cosmic rays, and magnetic
fields, yielding $\epsilon_B \la 0.2$.

It is quite possible that the winds in BAL quasars also carry a substantial
magnetic energy fraction, especially since some models predict that the BAL
wind is powered by magnetic forces (e.g. de Kool \& Begelman 1995).  For
illustrative purposes, we discuss a model with $f_{BAL}^{Pei} = 1.0$ and
$\epsilon_B = 0.01$; we refer to this scenario as the \emph{BALQSO model}.
The $\sim 10\%$ fraction of BAL systems among all quasars is
observationally found to be constant even at high redshifts
(Storrie-Lombardi et al. 2000; \cite{sdss}) and reflects the covering
fraction of their outflows in our model. The small value of $\epsilon_B$ is
chosen so that the outflows in this model may also characterize the vast
majority of all quasars that are radio quiet. We stress, however, that the
value adopted for $\epsilon_B$ is highly uncertain in this case.

We use the above two models to bracket the range of interest; together,
they allow us to discuss the physical implications of our scenario for
magnetic field generation in the IGM.  The RLQ model fills $\sim 20\%$ of
space with magnetized regions in which the magnetic and thermal energy of a
photo-ionized IGM are near equipartition.  Numerical simulations show that
the structure of the Ly$\alpha$ forest can be reconstructed without
including magnetic fields (Dav\'{e} \etal 1999).  Higher values of
$\epsilon_B$ may therefore conflict with existing data.  The BALQSO model
fills nearly all of the IGM with magnetized regions, and so we adopt a
small value of $\epsilon_B$ for it.

Figures 4-7 show results for the RLQ model.  Because the results scale
simply with $\epsilon_B$ and $f^{Pei}$, we do not show analogous plots for
the BALQSO model.  For ease of display, we define $\epsilon_{-1} \equiv
\epsilon_B/0.1$ and $\tilde{B} \equiv B/\sqrt{(\epsilon_B/0.1)}$.

\subsection{Filling Factor and Mean Magnetic Energy Density} 

Assuming that the sources follow a Poisson distribution in space, the
filling factor of their magnetized bubbles at redshift $z$ is $F(z) = 1 -
e^{-\phi(z)}$, where \bq \phi (z) = \int^{\infty}_{z} dz'
\int_{M_{h,min}}^{M_{h,max}} dM_h \frac{dn_q(M_h,z')}{dM_h dz'}
\hat{V}(z;M_h,z').
\label{eq:naivefillfactor}
\eq Here $\hat{V}(z;M_h,z')$ is the comoving volume at redshift $z$ of a
bubble produced by a quasar forming at redshift $z'$ in a host of mass
$M_h$.  The upper panel of Figure 4 shows the filling factor $F(z)$ for
various RLQ model scenarios.  The solid curves assume $f_d=0.8$ and that
$M_{h,min}$ is determined by atomic cooling considerations before
reionization and infall suppression afterward (top curve), $M_{h,min}=10^9
\msun$ (middle curve), and $M_{h,min} = 10^{10} \msun$ (bottom curve).  The
dashed curve assumes $f_d=0$ and that $M_{h,min}$ is determined by atomic
cooling and infall suppression.  All curves assume that
$f_R^{PS}=0.01$, $f_R^{Pei}=0.1$, $\tau_q=10^7 \yr$, and $\epsilon_B=0.1$.
In all cases, we find that the filling fraction is very small at
high-redshifts but rises to a substantial value at moderate and low
redshifts ($\sim 15$--$40\%$ at the present day).

It is straightforward to analytically approximate this integral for the low
redshift case of equation (\ref{eq:peinumber}).  Such an approximation
allows us to examine the dependence of the result on the model parameters.
For this estimate, we use equation (\ref{eq:radestimate}) for the bubble
radius and neglect the high and low halo mass cutoffs.  The result of the
integration is
\begin{eqnarray}
\phi_{Pei} (z) \approx 0.3 
\left( \frac{G(z,z_i)}{0.29} \right)
\left( \frac{f^{Pei}}{0.1} \right)
\left( \frac{\Phi_{\star}}{10^{4.34}\gpcden} \right) 
\left( \frac{\Gamma (32/5 - 4 \beta_{\star})}{4.48} \right)
\left[ \left( \frac{10^7\,\yr}{\tau_q} \right)
\left( \frac{0.7}{h} \right)
\right]^{2/5} \nonumber \\ 
\times \left[
\left(\frac{0.52}{K}\right) 
\left(\frac{\epsilon_k (1+ \epsilon_B)}{1.1} \right) 
\left( \frac{L_\star}{10^{9.78} \Lsun} \right)
\left( \frac{0.019}{\Omega_b h^2} \right)
\right]^{3/5}, \hskip 0.8 in
\label{eq:peiest}
\end{eqnarray}
where $G(z,z_i)$ is a function of the boundaries of the redshift interval
over which quasars are considered:
\bq
G(z,z_i) = \int_{z_i}^{z} dz' \frac{(1+z')^{1/5}}
{(\Omega_0 (1+z')^3 + (1 - \Omega_0))^{11/10}} \exp 
\left( \frac{-(z'-z_{\star})^2}{10 \sigma_{\star}^2/3} \right),
\label{eq:gdefn}
\eq and $G(0,4)=0.29$.  Quantities with a $\star$ subscript are
parameters from the Pei (1995) luminosity function: $\Phi_{\star}$ is
the space density normalization constant, $L_{\star}$ 
characterizes the `break' luminosity between the power law and
exponential portions of the fit, $-\beta_{\star}$ is the slope of the
power law portion, $z_{\star}$ is the peak of the quasar era, and
$\sigma_{\star}$ characterizes the duration of the quasar era.  (This
luminosity 
function was derived for an open model with $\Omega_0=0.2$ and no
cosmological constant.  We must leave it in this form in order to evaluate
the mass integral analytically, although our simulations do take place in a
flat universe with $\Omega_{\Lambda 0} = 0.7$.)

The analogous approximation for higher redshifts, when the source counts
are given by equation (\ref{eq:psnumber}), is less straightforward because
of the non-analytic form of the Press-Schechter mass function.  To begin,
we fit a power law, $\sigma(M) = B (M/\mbox{M}_{\odot})^{-\beta}$, to
the fluctuation 
spectrum over the range $M=10^8$--$10^{10} \msun$, which is the mass
interval most relevant for our calculation.  With our standard cosmological
parameters, the fluctuation spectrum is well-fit by $\beta = 0.0826$ and
$B=30.42$ in the range of interest.  Then a numerical integration of
(\ref{eq:naivefillfactor}), using equations (\ref{eq:radestimate}) and
(\ref{eq:psnumber}), yields
\begin{eqnarray}
\phi_{PS} (z=0) 
\approx 0.2
\left[
\left( \frac{0.52}{K} \right) 
\left(\frac{\epsilon_{bol}}{0.093}\right)
\left(\frac{\epsilon_{h}}{4 \times 10^{-4}} \right)
\left( \frac{\epsilon_K (1+\epsilon_B)}{1.1} \right)
\left( \frac{\tau_q}{10^7 \yr} \right)
\left( \frac{0.019}{\Omega_b h^2} \right)
\right]^{3/5}
\nonumber \\ \times
\left( \frac{f^{PS}}{0.01} \right) 
\left( \frac{\Omega_0}{0.3} \right) 
\left( \frac{h}{0.7} \right)^{4/5}, \hskip 2.8 in 
\label{eq:psest}
\end{eqnarray} 
where the dependence on $\Omega_0$ comes from the linear growth factor.  
(We have neglected the dependence of the growth factor on the
cosmological constant; however, its effects will be negligible for the
quasars at high redshifts to which we apply the formula.)
In this estimate we assume that the minimum halo mass is determined by 
cooling before reionization and infall suppression afterward.
Note the different dependence on $f$ and $\tau_q$
between this expression and the low-redshift result, equation
(\ref{eq:peiest}): $\phi_{Pei} \propto f^{Pei} \tau_q^{-2/5}$ while $\phi_{PS}
 \propto f^{PS} \tau_q^{3/5}$.  
This is a direct result of the matching procedure between the two
source counts described by equation (\ref{eq:matchcounts}).

Figure 5 shows the relative contributions of quasars from the high and low
redshift regimes for the RLQ model, calculated with our full numerical
model.  The upper panel shows the filling
fraction $F(z)$ for quasars with source counts determined by equation
(\ref{eq:psnumber}) at $z_0 > 4$ (solid line) and by equation
(\ref{eq:peinumber}) at $z_0 < 4$ (dashed line).  Both curves assume
$f_R^{PS} = 0.01$, $f_R^{Pei} = 0.1$, $f_d = 0.8$, $\tau_q=10^7 \yr$,
$\epsilon_B = 0.1$ and $z_r = 7$.  As suggested by the estimates
(\ref{eq:peiest}) and (\ref{eq:psest}), the contributions to the filling
factor at $z=0$ are comparable for the two regimes.  Note that for $z \lesssim
4$, the filling factor from high-$z$ sources is approximately constant with
time, indicating that most of the bubble remnants from this population have
reached their maximum comoving radii by this point. Equations
(\ref{eq:peiest}) and (\ref{eq:psest}) are accurate to 
within a factor of $\sim 1.5$, despite the fact that 
they do not include proper cosmological parameters or mass cutoffs and
that they ignore the detailed dynamics of the expanding bubbles.

Estimates (\ref{eq:peiest}) and (\ref{eq:psest}) imply that the dependence
of the filling factor $F(z)$ on the initial magnetic energy fraction is
very weak, as $\phi \propto (1+\epsilon_B)^{3/5}$.  The BALQSO model, which
assumes that every quasar hosts a BAL outflow, is therefore more efficient
at polluting large volumes than the RLQ model.  An imposed low-mass cutoff
can dramatically decrease the contribution from high-$z$ sources, because
massive halos are considerably rarer at these epochs.  However, it has
little effect on the low-$z$ contribution, because the nonlinear mass scale
at these redshifts is well above the low-mass cutoff.

Equations (\ref{eq:peiest}) and (\ref{eq:psest}) ignore the high-mass
cutoff at $M_{esc}$.  
Because this limiting mass is a
fairly strong function of the model parameters [for example, $M_{esc}
\propto \tau_q^{3/2}$; see equation (\ref{eq:massest})], changing
these parameters can sometimes have a 
dramatic effect on the results.  This is particularly true at high-$z$,
when the minimum mass for efficient atomic cooling and the high-mass cutoff
are close.  For example, at $z=15$, the minimum mass for cooling is $\sim
10^8 \msun$.  If $\tau_q = 10^6 \yr$ and $f_d=0.8$, then $M_{esc} \sim 5
\times 10^9 \msun$ at this redshift, while if $\tau_q = 10^7 \yr$ and
$f_d=0.8$ then $M_{esc} \sim 3 \times 10^{11} \msun$.  
The shrinking of the allowed halo mass interval in the former case
means that short 
lifetime quasars are less efficient (by a factor $\sim 1.25$) than we would
expect from the predicted scaling $\phi_{PS} \propto f^{PS}
\tau_q^{3/5}$ of equation (\ref{eq:psest}); the resulting
$z=0$ filling fraction from high-redshift sources  in this scenario
is $F_{PS} \approx 0.3$.  

The global volume-averaged magnetic energy density is 
\bq
\bar{u}_B(z) = \int^{\infty}_{z} dz' \int_{M_{h,min}}^{M_{h,max}}
dM_h \frac{dn_q(M_h,z')}{dM_h dz'} \hat{V}(z;M_h,z') u_B(z;M_h,z'),
\label{eq:avgmagpress}
\eq
where $u_B(z;M_h,z')$ is the mean magnetic energy density at redshift
$z$ inside a bubble produced by a quasar forming at redshift $z'$ in a
host of mass $M_h$.  

The lower panel of Figure 4 shows the ratio between
$\bar{u}_B(z)/\epsilon_{-1}$ and a fiducial value for the mean thermal
energy density of the IGM, $u_{fid} = 3 n(z) kT_{IGM}$, where $T_{IGM} =
10^4 \kel$ and $n(z)$ is the mean cosmic baryon density, for the same
scenarios as shown in the upper panel of Figure 4.  We find that this ratio is
approximately independent of $\epsilon_{-1}$.  The above fiducial energy
density was chosen only for normalization purposes. It corresponds to a
uniform IGM that has been heated by a photo-ionizing radiation field, and
hence it grossly overestimates the typical pressure in the neutral IGM
before reionization $z\ga 7$--$10$ (Gnedin \& Ostriker 1997; Gnedin 2000) and
underestimates the temperature that the IGM acquires in large-scale shocks
at low redshifts $z\la 3$ (Cen \& Ostriker 1999; Dav\'e et al. 1999).  The
signature of the `quasar era' is apparent in the figure; once the quasar
abundance turns over, the mean magnetic energy density decreases as well.

Because the bubble size is nearly independent of the injected magnetic
energy [equation (\ref{eq:radestimate})], the mean magnetic energy density
of each bubble is approximately proportional to $\epsilon_B$.  The global
energy density is then proportional to that of a single bubble
multiplied by the global filling factor, so that $\bar{u}_B \propto
\epsilon_B F(z) \propto \epsilon_B f^{Pei}$.  Therefore the \emph{global}
average energy density is approximately the same in the RLQ and BALQSO
models: the increased number of sources in the BALQSO model
compensates for the 
weaker magnetic fields within the bubbles of that model.  Nevertheless, the
local dynamics are quite different.  For the RLQ model, we find that
$\bar{u}_B/u_{fid} \approx F(z)$ at $z \lesssim 7$.  This indicates that,
in regions filled by magnetized bubbles, the magnetic and thermal pressures
have reached approximate equipartition in this model.  On the other hand,
the BALQSO model has $\bar{u}_B/u_{fid} \approx 0.1 F(z)$, so that thermal
pressure still dominates inside and around the bubbles.

The lower panel of Figure 5 shows the separate contributions to the ratio
between the volume-normalized magnetic energy density
$\bar{u}_B/[\epsilon_{-1} F(z)]$ and $u_{fid}$ for the RLQ model by quasars
at $z_0 > 4$ (solid line) and at $z_0 < 4$ (dashed line) with the same
parameter choices as in the upper panel.  In each case, we divide the
magnetic energy density by the appropriate filling factor in order to
display the magnitude of the magnetic energy in the bubbles, rather than
the global volume average.  As expected, the importance of the magnetic
energy within bubbles declines with redshift.  At $z=3$ we find that $\sim
15\%$ of space is filled with magnetic fields that are, on average,
slightly below equipartition, while a much smaller fraction of space ($\sim
0.3\%$) contains strong fields.  The remnants of high-redshift quasars fill
approximately $20\%$ of space at $z=0$; in these regions, the average
magnetic field strength corresponds to $u_B \sim 0.03 u_{fid}$.  The much
younger remnants of low-redshift quasars fill $\sim 20\%$ of space with
stronger field regions (where $u_B \sim 0.8 u_{fid}$, on average).  We
stress that these magnetic energy values are averages.  More details on the
distribution of field strengths at different epochs are given in the
following section.

As mentioned in \S 2.1.4, our model underestimates the ambient pressure at
low redshifts, because many sources will be embedded within galaxy groups
or clusters.  The gas in such groups is not only at a higher density than
the diffuse IGM, but it also has a higher temperature (approximately the
virial temperature of the group or cluster).  Figure 5 shows that even if
the higher pressure in groups can suppress the expansion of bubbles around
low-$z$ quasars, a high-$z$ component in the RLQ model will
nevertheless pollute $\sim 20\%$ of space with low-level fields.  Of
course, the enhanced pressure inside groups will not halt the
formation of quasars.  Therefore, magnetic energy will 
still be expelled into the intragroup medium and possibly get mixed by
motions of galaxies or mergers with other groups.  The result may be that
the groups collapse with a low-level field, which subsequent outflows
enrich by injecting magnetic energy directly into the intragroup or
intracluster 
medium.  The net result would be little different from the picture outlined
above, except that the high-field regions created by low-$z$ quasars would
be strongly correlated with groups and clusters.

\subsection{Bubble Size and Magnetic Field Distribution}

Next we would like to find the probability distribution functions of the
bubble magnetic field and radius. Let $P(B,z)dB$ be the probability at
redshift $z$ that a random point in space is contained within a bubble
having $<$$B^2$$>^{1/2}$ in the range $(B,B+dB)$.  The probability at
redshift $z$ that the bubble was produced by a quasar that formed at $z'$
is \bq P(B,z,z')dB \propto \hat{V}(z,M_h,z') \left(\frac{dn_q}{dM_h dz'}
\right) \frac{d M_h(B,z,z')}{d B} dB.
\label{eq:pbzzdefn}
\eq 
Then, $P(B,z)dB \propto dB \int_z^{\infty} P(B,z,z') dz'$. As
described in \S 3.1, the magnetic energy density 
inside a bubble scales linearly with $\epsilon_B$; therefore,
we discuss $P(\tilde{B}, z)$, which is approximately independent of
$\epsilon_B$. 

Figure 6 presents the probability distribution $P(\tilde{B},z)$ of
magnetic fields for the  
RLQ model at a series of redshifts ($z=0,3,6,9,12$, and $15$, from
left to right).  All curves are normalized to have a
total area of unity and assume $f_R^{PS} = 0.01$, $f_R^{Pei} = 0.1$,
$f_d=0.8$, and $\tau_q=10^7 \yr$.  
The pronounced asymmetry in the distribution function at low redshifts is
due to the substantial increase in the minimum halo mass after
reionization. The rise at high $B$ in the $z=0$ curve is because
low-redshift halos tend to be large.
Magnetic flux conservation and equation (\ref{eq:radestimate}) imply that
$B \propto M_h^{1/5}$, if the quasar redshift is held constant.  The quasar
era produces a surplus of massive halos, and the corresponding bias to
large magnetic fields can be clearly seen by the dotted line in Figure 6,
which shows the contribution of quasars with $z_0 < 4$ to
$P(\tilde{B},0)$.  The 
dashed line shows the contribution from quasars at high redshifts ($z_0 >
4$); these make up the low-field regions.

The probability distribution of comoving bubble radii, $P(\hat{R},z)$, is
calculated in an exactly analogous fashion and is shown in Figure 7 for the
same series of redshifts ($z=15,12,9,6,3$ and $0$, from left to right) and
same parameter choices as in Figure 6. This distribution function is
approximately independent of $\epsilon_B$ [by equation
(\ref{eq:radestimate}), $\hat{R}_{max} \propto (1 + \epsilon_B)^{1/5}$].
The similarity in trends between the two figures is clear.  The dotted
curve again shows the contribution to the $z=0$ distribution function from
quasars with $z_0 < 4$ while the dashed curve shows the contribution from
$z_0 > 4$ quasars.  Note that the low-$z$ distribution is strongly biased
toward large bubbles, because the contributing halos are massive.

Because the probability distribution of bubble radii depends only weakly on
$\epsilon_B$ while the peak of the magnetic field distribution scales
approximately as $B_{peak} \propto \sqrt{\epsilon_B}$, the BALQSO model
produces bubbles with characteristic sizes very similar to those of the RLQ
model, with each bubble containing weaker magnetic fields.  A minimum-mass
cutoff shifts the peak to a larger radius and magnetic field strength,
because higher-mass halos generically produce larger bubbles with stronger
magnetic fields.  Setting $f_d=0$ shifts $P(\hat{R},z)$ to smaller
radii (with mean size $\sim 1 \mpc$ at $z=0$) 
but also shifts $P(B,z)$ toward stronger magnetic fields (with a mean
at $B \sim 2 \times 10^{-9} \gauss$), because a bubble with a
given magnetic energy input is physically smaller, and hence its fields are
less diluted.

\subsection{Reionization by Quasar Outflows}

This model, with minor modifications, also allows us to examine the
feasibility of cosmological reionization by quasar outflows.  We use an
approach similar to that of Tegmark, Silk, \& Evrard (1993), who examined
collisional reionization due to starburst winds from low-mass galaxies
in the early universe.

In this case, we must track the ionization history of each bubble.  A
simple way to do so is to follow the temperature of the interior of each
bubble [equation (\ref{eq:bubbletemp})].  As described in \S 2.1.3, if $T_b >
1.5 \times 10^4 \kel$, the hot gas in the bubble will ionize a
fraction $f_m ~ 0.1$ of the incoming hydrogen; if the
bubble temperature is below this critical value, incoming hydrogen will
remain neutral.  An estimate of the ionized \emph{volume} produced by
each bubble 
is therefore its comoving volume when this critical temperature is reached,
$\hat{V}_T$.  Later, the ionized hydrogen may recombine on a timescale
$(\alpha_B n_b)^{-1}$, where $\alpha_B$ is the case-B
recombination coefficient for hydrogen.  We find that, because of the
small internal density of the bubbles, very few bubbles
recombine before $z \sim 5$.  Recent observations of high-$z$ quasars
show that reionization 
occurred at $z \ga 6$ (\cite{fan}), and so we neglect the effects of
recombination in the bubbles.

Given the ionized volume around each quasar, we can now find the filling
factor of ionized material, $F_r(z)$, in an analogous manner to the filling
factor of the magnetized bubbles; we need only replace $\hat{V}$ by
$\hat{V}_T$ in equation (\ref{eq:naivefillfactor}).  In this calculation,
the minimum halo mass is determined by the condition for atomic cooling.
We include quasars at $z_0>4$ only, so the source counts are determined by
the Press-Schechter formalism through equation (\ref{eq:psnumber}).

Note that an estimate of the ionized \emph{mass fraction} would be very
different than that calculated here.  A fraction $(1-f_m) \sim 0.9$ of the
material in the blastwave is contained in the shell, which cools much
faster than the hot bubble cavity and, because of its higher density, will
also recombine faster than the cavity.  Therefore, our scenario will be
considerably more efficient at ionizing a large volume of space than at
reionizing a large fraction of the hydrogen atoms.

As shown by equation (\ref{eq:radestimate}), $\epsilon_B$ has only a small
effect on the final radius of the bubbles, so collisional reionization is
nearly independent of the magnetic field strength.  The BALQSO model is
thus most efficient at reionization (because $f_{BAL} = 10 f_R$), and so we
focus on that model in this section.

The reionization filling factor, $F_r(z)$, is plotted in Figure 8.  For
illustrative purposes, we show results for the two extreme cases, $f_d=1$
(solid line) and $f_d=0$ (short dashed line); each of these assumes that
$\tau_q=10^7 \yr$, $f_{BAL}^{PS} = 0.1$, and $\epsilon_B = 0.1$ .  Also
shown are cases in which $\tau_q=10^6 \yr$, $f_{BAL}^{PS} = 1.0$, and
$\epsilon_B=0.1$, with $f_d=1$ (long dashed line) and $f_d=0$ (dot-dashed
line).  [The last two models illustrate different ways to match the high-
and low-$z$ source counts; see the discussion accompanying equation
(\ref{eq:matchcounts}).]

Clearly, only in the most optimistic scenarios is complete reionization by
quasar outflows possible.  Very high rates of injection of dissipated
energy into the bubble interiors ($f_d \sim 1$) are required in order to
reionize even $\sim 80\%$ of space before the existing observational limit
($z \sim 6$).  More conservative scenarios predict that only $\sim
50\%$ of space can be ionized by the quasar bubbles even at the present
day.  By this point, recombination inside the bubbles has further reduced
the filling factor by $\sim 30\%$.  We conclude that collisional ionization
by quasar ouflows is a less viable scenario for the reionization of the IGM
than is photo-ionization (see, e.g. Gnedin 2000).

\section{Discussion}

We predict that the intergalactic medium (IGM) has a cellular magnetic
structure, with highly magnetized bubbles representing the old remnants of
quasar outflows.  Different types of outflows result in different outcomes
in our models.  Radio-loud quasars at low redshifts ($\tau_q = 10^7 \yr$,
$\epsilon_B \gtrsim 0.1$, $f_R^{Pei} \gtrsim 0.1$, and $z_0 < 4$) generate
high field regions that fill $\sim 20\%$ of the photo-ionized IGM volume
with regions in which the magnetic and thermal energy densities are 
near equipartition. An (as yet unobserved) population of high
redshift radio-loud quasars ($\tau_q = 10^7 \yr$, $\epsilon_B \gtrsim 0.1$,
$f_R^{PS} \gtrsim 0.01$, and $z_0 > 4$) fill a further $\sim 20\%$ of space
with bubbles in which the magnetic energy density is $\sim 3\%$ of the
thermal energy density.  However, the potentially less magnetized outflows
from broad-absorption-line quasars ($\tau_q = 10^7 \yr$, $\epsilon_B
\gtrsim 0.01$, $f_{BAL}^{Pei} \sim 1$, and $f_{BAL}^{PS} \gtrsim 0.1$) may
fill most of space with a weaker magnetic field.
In reality, a combination of these processes will likely have produced a
low-level field filling most of space, with high-field peaks generated by
radio-loud quasar activity.

As shown by Figure 5, sources at both high and low redshifts can have
substantial effects on the intergalactic magnetic field (IGMF); our
model predicts that the contribution 
to the total filling factor of quasars at $z_0 < 4$ (which we describe using
the observed luminosity function) and those at $z_0 > 4$ (which
we describe through the Press-Schechter mass function) are comparable.
However, because the high-$z$ sources reside in less massive halos and are
therefore generically less luminous and because the bubbles produced by these
sources have expanded for longer times, the volume filled by the high-$z$
sources has a considerably weaker magnetic field than that filled by the
observed low-$z$ population.

Our model provides only an upper limit to the coherence length of the
IGMF, because we do not follow the field structure inside each bubble.
An observational limit on this coherence length can be obtained from the
mean Faraday rotation signal of distant sources.  To estimate an upper
limit to this effect from our model, we assume that the field within each
bubble is coherent over the entire bubble diameter and that all bubbles at a
given redshift are identical, with the characteristic size $r(z)$ given by
the mean of the probability distribution of bubble radii.  The number
density of bubbles at a redshift $z$ is then $N(z) = 3 \phi(z)/[4 \pi
r^3(z)]$, where we use $\phi(z)$ rather than $F(z)$ in order to include the
possibility of overlapping bubbles.  The probability per unit path-length
of encountering a bubble centered at $z$ is $\pi N(z) r^2(z)$, so the
expected squared rotation angle $d \varphi^2$ from bubbles in the redshift
interval $(z,z+dz)$ is 
\bq d \varphi^2 = \left( \frac{4 r(z)}{3}
\frac{e^3}{2 \pi m_e^2 c^2 \nu^2} n_e(z) \frac{B(z)}{\sqrt{2}}
\right)^2 \pi N(z) r^2(z) 
\left| \frac{c dt}{dz} \right| dz,
\label{eq:dphi}
\eq where $n_e(z)$ is the electron number density, $m_e$ is the electron
mass, $e$ is the electron charge, $\nu_0 = \nu/(1+z)$ is the observed
frequency, and $B(z)$ is the 
redshift-dependent mean bubble magnetic field strength.  We have
assumed a random impact parameter through each bubble and 
that the magnetic field of each bubble is randomly oriented with respect to
the line of sight.
Faraday rotation results are usually quoted in terms of the 
rotation measure, RM$= \Delta (<\varphi^2>^{1/2}) / \lambda_0^2$, 
where $\lambda_0=c/\nu_0$.  
For our standard RLQ and BALQSO scenarios, RM$(z=20) \sim 0.7
\radmsq$ and RM$(z=2.5) \sim 0.5 \radmsq$.  Observations limit RM$(z=2.5)
\lesssim 5 \radmsq$ (\cite{kron94}), an order-of-magnitude above our upper
limit.  Thus, the coherence length of the field could easily stretch across
the entire scale of each bubble without violating existing constraints.

Interestingly, an early generation of quasars will have left magnetized
remnants in the present-day IGM.  It may be possible to observe these
`fossil bubbles' through two techniques.  The most direct method would be
to observe Faraday rotation of a background source through the bubbles.
Unfortunately, the rotation through a single bubble would be extremely
weak, even assuming a coherent field across the entire bubble oriented
along the line of sight (RM$\,\sim 3 \times 10^{-4} \radmsq$ for a typical
bubble at $z=0$ with $R \sim 1 \mpc$ and $B \sim 10^{-9} \gauss$).
Alternatively, it may be possible to observe the
synchrotron halos produced by these bubbles, if some of them contain a
population of relativistic electrons.  Such electrons could be produced in
shocks in the IGM (see below) or by compression of the radio plasma during
the formation of galaxy clusters (\cite{ensslin}).  In fact, `diffuse radio
halos' with sizes of $\sim 1 \mpc$ are observed around the cores of many
X-ray clusters, and `radio relics,' somewhat smaller radio halos, are often
observed near the periphery of clusters (\cite{feretti}).  Both of these
objects are thought to be remnants of powerful radio sources embedded in
the cluster (\cite{ensslin}); the observed correlation of the radio relics
and halos with merging clusters (\cite{feretti}) suggests that shock waves
from structure formation can accelerate electrons to relativistic energies
and thus `turn on' the synchrotron halos.  Similar processes would occur
for our bubbles, although the low densities and magnetic field strengths in
the uncompressed IGM imply that the halos have only a very low surface
brightness (Waxman \& Loeb 2000).

The implied magnetic fields should lead to the acceleration of electrons
and protons to relativistic energies in the strong intergalactic shocks
that are produced by converging flows as large-scale structure forms in the
IGM. The inverse-Compton scattering of the microwave background by the
accelerated electrons could account for a substantial fraction of the
$\gamma$--ray background (Loeb \& Waxman 2000) and would also be
accompanied by radio synchrotron emission due to the same electrons (Waxman \&
Loeb 2000). Observations of the $\gamma$-ray and radio emission in the
brightest shocks around clusters of galaxies can be used to calibrate the
magnetic field strength in these regions and to test our model. Unlike Faraday
rotation measurements, this approach is not sensitive to the coherence
length of the magnetic field (since the typical Larmor radius of the
electrons is negligible, $\la 1$ pc). The future GLAST mission\footnote{See
http://www-glast.stanford.edu/} will have the sensitivity and angular
resolution necessary to map the predicted $\gamma$-ray luminosity of galaxy
clusters and correlate it with low-frequency radio maps of the same regions
(Loeb \& Waxman 2000; Waxman \& Loeb 2000).

As the IGM gas collapses to form bound systems, such as galaxies and
clusters of galaxies, the magnetic field is compressed further
adiabatically with $u_{B}=B^2/8\pi \propto \rho_g^{4/3}$. The thermal
energy density increases even faster in the adiabatic regime, $u_T\propto
\rho_g^{5/3}$, and could be enhanced further as the specific entropy
of the gas is
increased at the virialization shock around these collapsed systems. Hence,
as long as gas cooling is unimportant (e.g. outside the core of X-ray
clusters), the magnetic field is expected to make only a small contribution
to the total pressure.  The typical values for our predicted IGMF at the
mean IGM density translate to a mixed magnetic field strength of $\sim 0.15
{\sqrt\epsilon_{-1}} \microgauss$ ($\sim 1.5 {\sqrt\epsilon_{-1}}
\microgauss$) at an overdensity of $\sim 10^3$ in the RLQ (BALQSO)
model, comparable 
to the inferred field amplitude on Mpc scales around X-ray clusters
(Fusco-Femiano et al.  1999; Rephaeli et al. 1999; Kaastra et al. 1999; Kim
et al. 1989).  The most important unknown parameter in
our model is $\epsilon_B$, the ratio of magnetic to kinetic energy in
quasar outflows.  Shear flows during cluster collapse could enhance the
IGMF (\cite{dolag}) and compensate for low values of $\epsilon_B$.

This model for the formation of the IGMF has several advantages over other
scenarios.  The fields are produced and amplified in the accretion flow
around a central black hole, for which the amplification time is short
(Balbus \& Hawley 1991; \cite{colgate-li}), so that fields in high-$z$
galaxies (Oren \& Wolfe 1995) 
may be easily accomodated.  Compared to a cosmological origin in
the early universe, the fields are produced sufficiently late that
causality does not limit the comoving coherence length.  Unlike the
scenario of Kulsrud \etal (1997), there is a natural large-scale coherence
length in this model, the characteristic bubble size at each redshift.
However, note that we do not follow the geometry of the magnetic field, so
the actual coherence length may be smaller.

Kronberg \etal (1999) proposed that winds from starbursting dwarf galaxies
in the early universe could carry the magnetic fields generated in stars
into the IGM.  Similarly to our model, they find that an early population
of sources can easily pollute $\gtrsim 10\%$ of space with magnetic 
fields.  A
starburst model for magnetic field generation, with a large number of
sources having relatively small magnetic 
energy fractions, would be most
similar to our BALQSO model, although individual magnetized bubbles are
likely to be smaller in a starburst model.  However, the coherence length
of the field should be extremely small in this case due to the large number
of stars contributing to each bubble (in contrast to the single source in
our model). Kronberg \etal (1999) are forced to make recourse to
amplification processes during the wind expansion in order to create a
coherent field on subgalactic scales ($\sim 8 \kpc$, the size of the wind
bubble in the nearby starburst galaxy M82) and subsequent `acausal
diffusion' in order to create larger coherent fields.  

Finally, we also found that mechanical outflows from quasars most likely
cannot reionize the universe before the observed limit, $z_r \gtrsim 6$,
even under the most favorable circumstances.  Photo-ionization by stars or
quasars is therefore required.

The implementation of our quasar outflow model into a numerical simulation
of the IGM can be used to explore the spectroscopic fingerprints that the
magnetic bubbles leave on the Ly$\alpha$ forest.  Bryan \etal (1999)
and Theuns \etal (1998) have found that existing simulations of the
Ly$\alpha$ forest
underestimate the median Doppler width of the absorption lines by
$\sim 50\%$ as 
compared to the observed median width, despite the close agreement in
simulated and observed column density distributions.  To date,
numerous explanations have been offered, most postulating extra heat sources
for the IGM not included in the simulations (Cen \& Bryan 2000, and
references therein).  In our model,
the IGMF will provide a non-thermal contribution to the pressure of
the gas.  An equipartition field will increase the line width by a
factor of $\sqrt{2}$, which would account for the claimed
discrepancy.  Such regions fill $\sim 20\%$ of space at $z=3$ in our
model, suggesting that nonthermal pressure from magnetic fields may
play a significant role in determining these line widths.

\acknowledgements 

We thank Rennan Barkana and Martin Rees for helpful
comments on the manuscript.  
This work was supported in part by NASA grants NAG 5-7039, 5-7768, and NSF
grants AST-9900877, AST-0071019 for AL.  SRF acknowledges the support
of an NSF graduate fellowship.

\begin{figure} [t]
\plotone{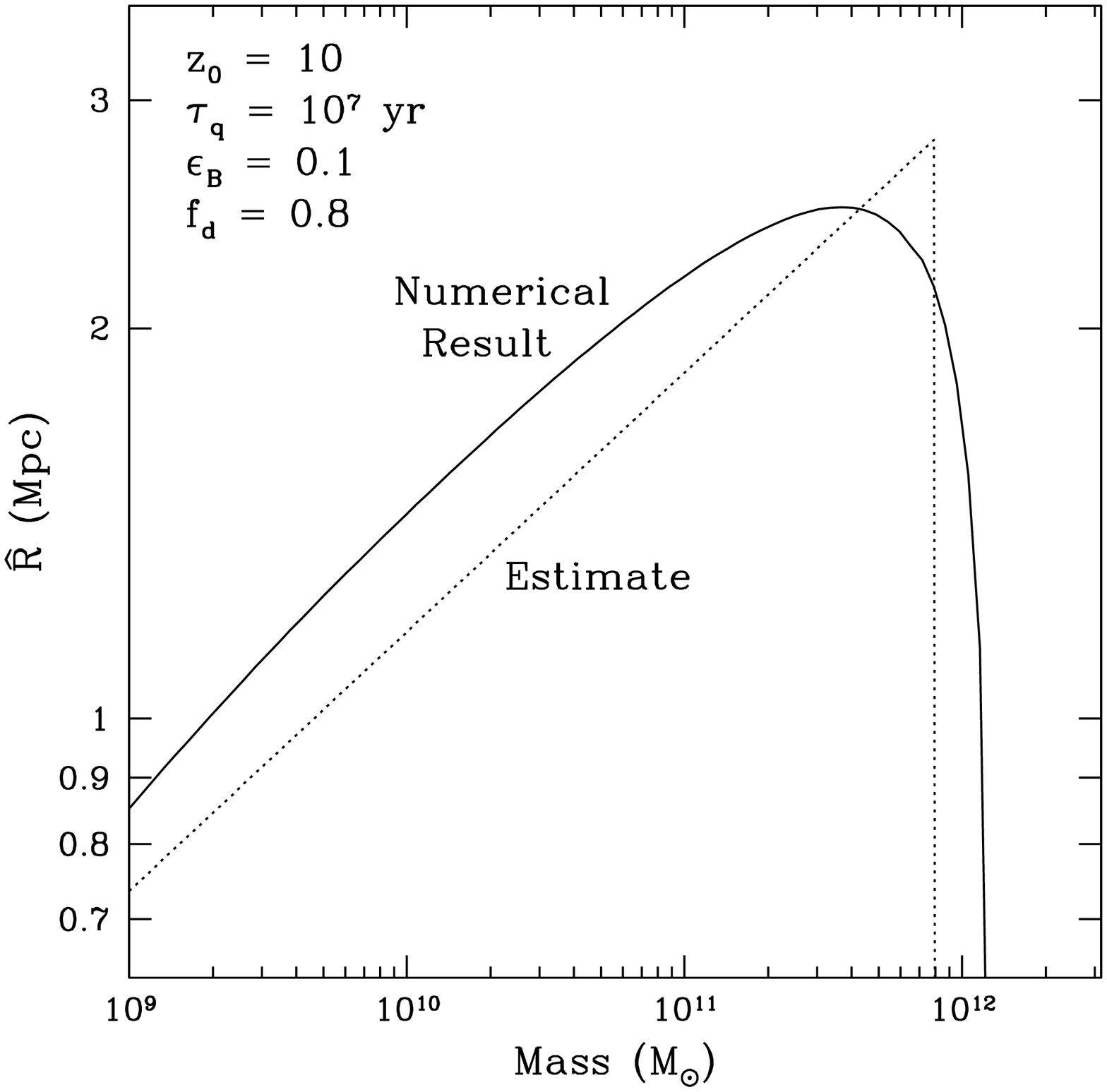}
\caption{Final ($z=0$) bubble radius $\hat{R}$ as a function of halo
mass.  The solid line shows the results of our numerical model with
$f_d=0.8$, while
the dotted line shows the estimate of equation
(\ref{eq:radestimate}) with $K=0.52$ and a mass cutoff determined by equation
(\ref{eq:massest}).  Both curves assume standard parameter values,
with $z_0=10$, $\tau_q = 10^7 \yr$, and $\epsilon_B = 0.1$.
}
\end{figure}

\begin{figure} [t]
\plotone{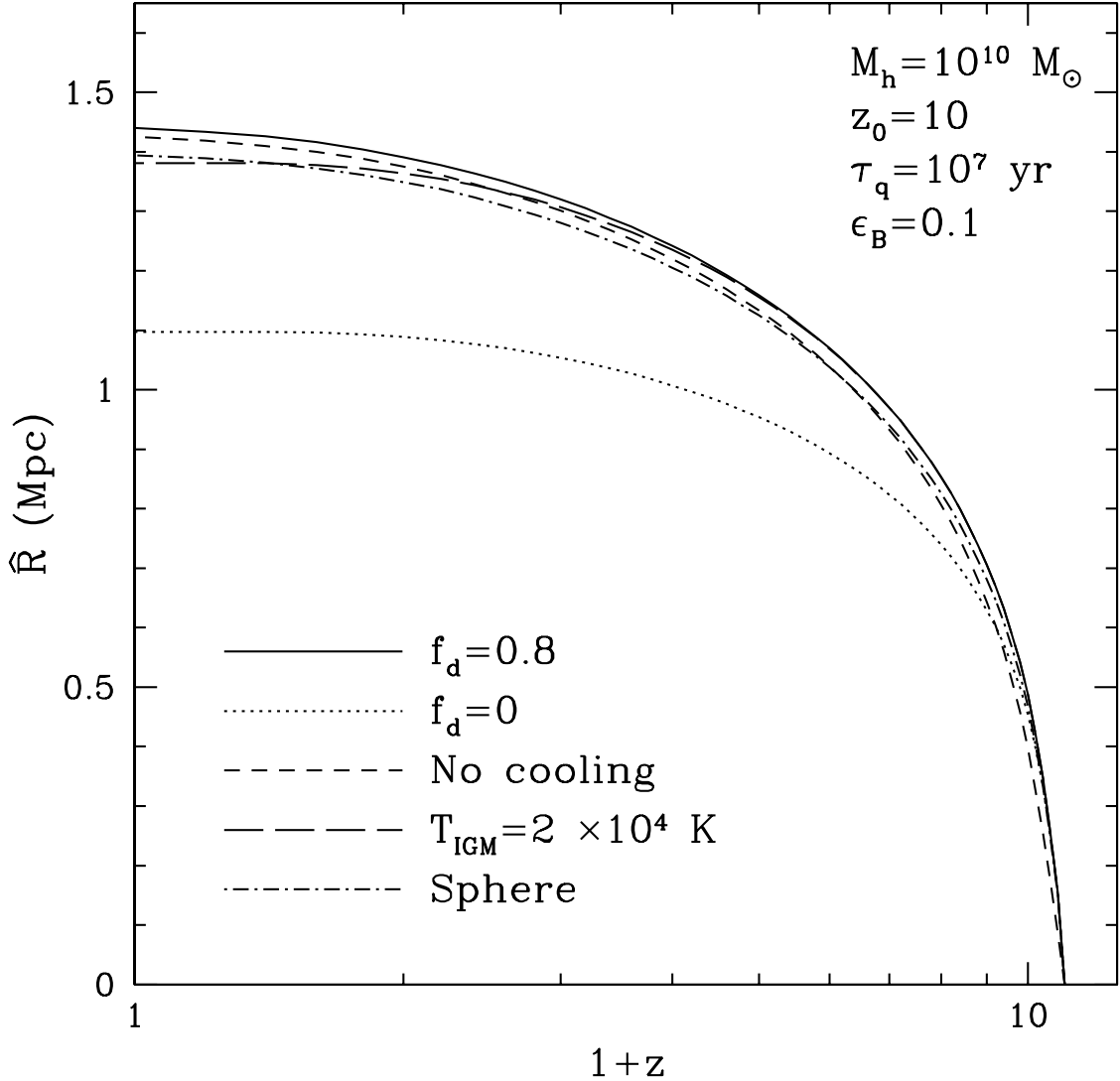}
\caption{Comoving bubble radius $\hat{R}$ as a function of redshift for a
variety of scenarios.  All cases assume $M_h = 10^{10} \msun$, $z_0 =
10$, $\tau_q=10^7 \yr$, and $\epsilon_B=0.1$.
Results are shown for our standard scenario
(solid line), a case in which the gas in the halo of the host galaxy
does not cool (short dashed line), a case in which
$T_{IGM} = 2 \times 10^4
\kel$ after reionization (long dashed line), and a case in which the
outflow is 
initially spherical rather than a jet (dot-dashed line); all of these
assume $f_d=0.8$.  Also shown is a case in which $f_d=0$ (dotted
line).
}
\end{figure}

\begin{figure} [t]
\plotone{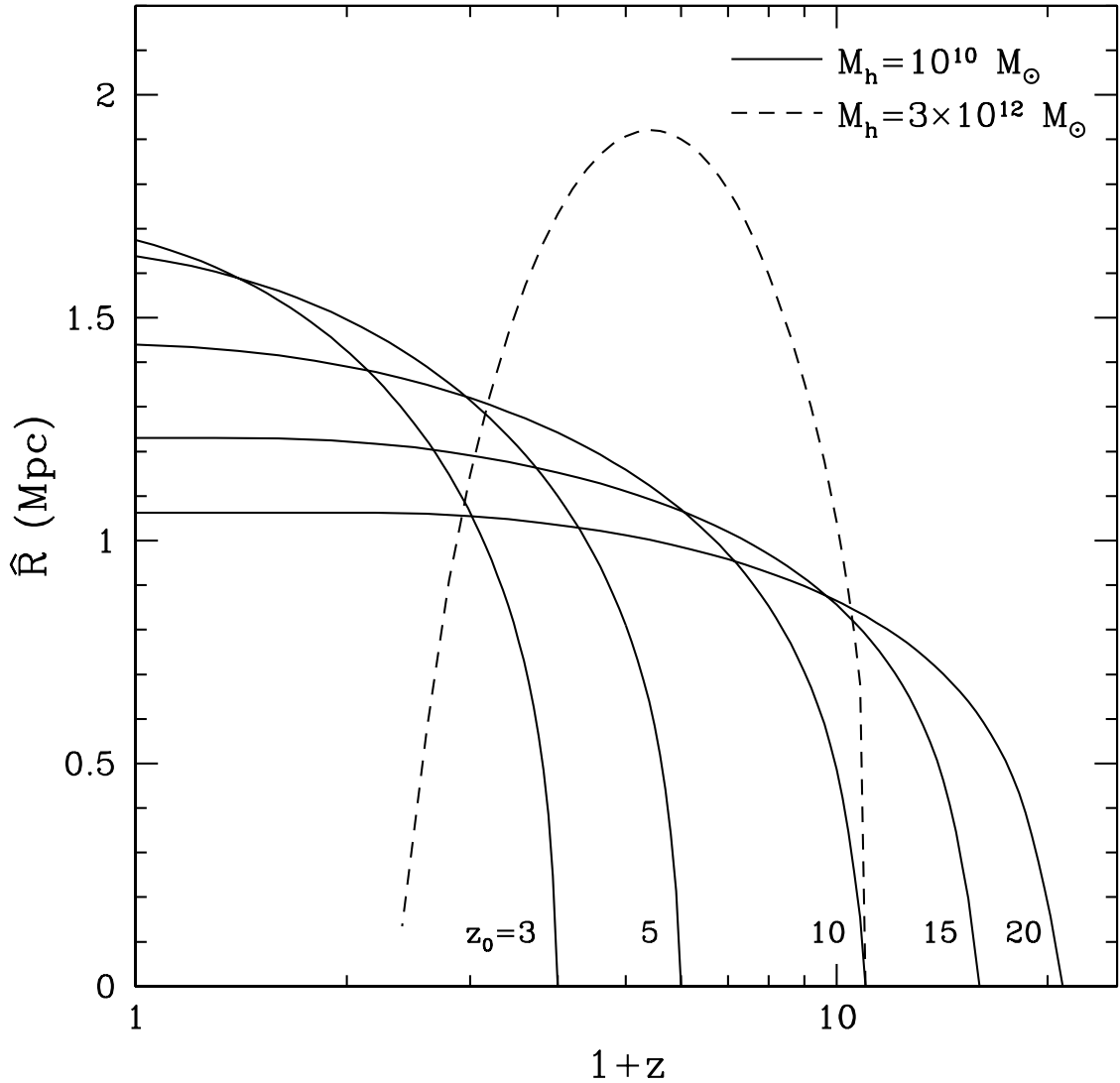}
\caption{ Comoving bubble radius $\hat{R}$ as a function of redshift for
quasars which produce an outflow at different redshifts, $z_0$.  For the
solid curves, $M_h = 10^{10} \msun$.  Beginning from right to left, $z_0 =
20,15,10,5, \mbox{ and } 3$.  The dashed curve shows the radius for $M_h =
3 \times 10^{12} \msun$ and $z_0 = 10$, and illustrates the recollapse
of a bubble 
for a massive halo in which gravity prevents escape into the
IGM.  In all cases, $\tau_q=10^7 \yr$, $\epsilon_B=0.1$, and $f_d=0.8$. }
\end{figure}

\begin{figure} [t]
\plotone{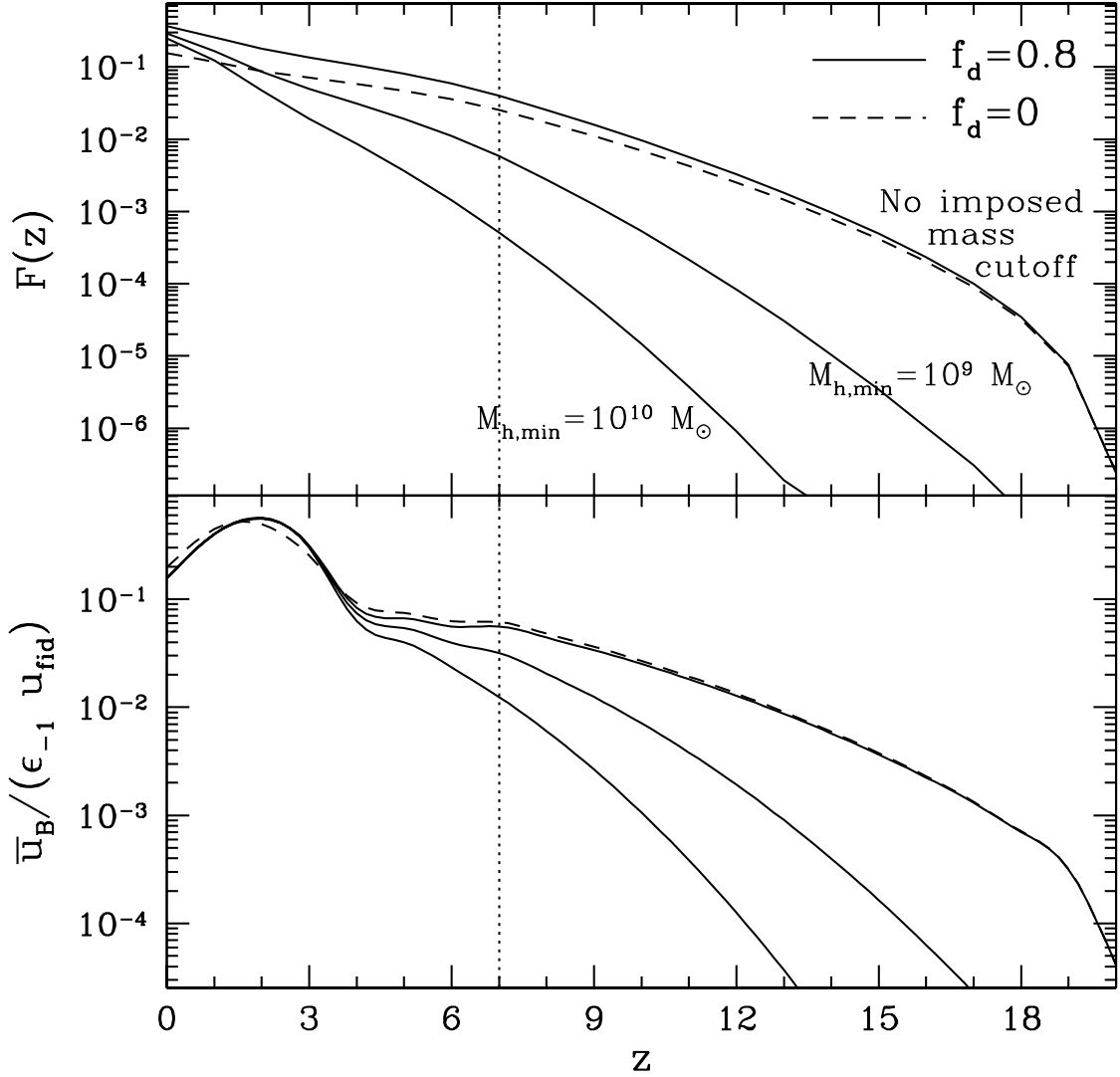}
\caption{\emph{Upper Panel:} Volume filling fraction of magnetized bubbles
$F(z)$, as a function of redshift, for the RLQ model.  \emph{Lower Panel:}
Ratio of normalized magnetic energy density, $\bar{u}_B/\epsilon_{-1}$, to
the fiducial thermal energy density $u_{fid} = 3 n(z) k T_{IGM}$, where
$T_{IGM} = 10^4 \kel$, as a function of redshift.  In each panel, the solid
curves assume $f_d=0.8$ and that $M_{h,min}$ is determined by atomic cooling
before reionization and infall suppression afterward (top curve), $M_{h,min}
= 10^9 \msun$ (middle curve), and $M_{h,min} = 10^{10} \msun$ (bottom
curve). The dashed curve assumes $f_d = 0$ and determines $M_{h,min}$ by
atomic cooling and infall suppression.  The vertical dotted line
indicates the assumed redshift of reionization, $z_r=7$. All 
curves assume $f_R^{PS}=0.01$, $f_R^{Pei} = 0.1$, and $\tau_q=10^7 \yr$. }
\end{figure}

\begin{figure} [t]
\plotone{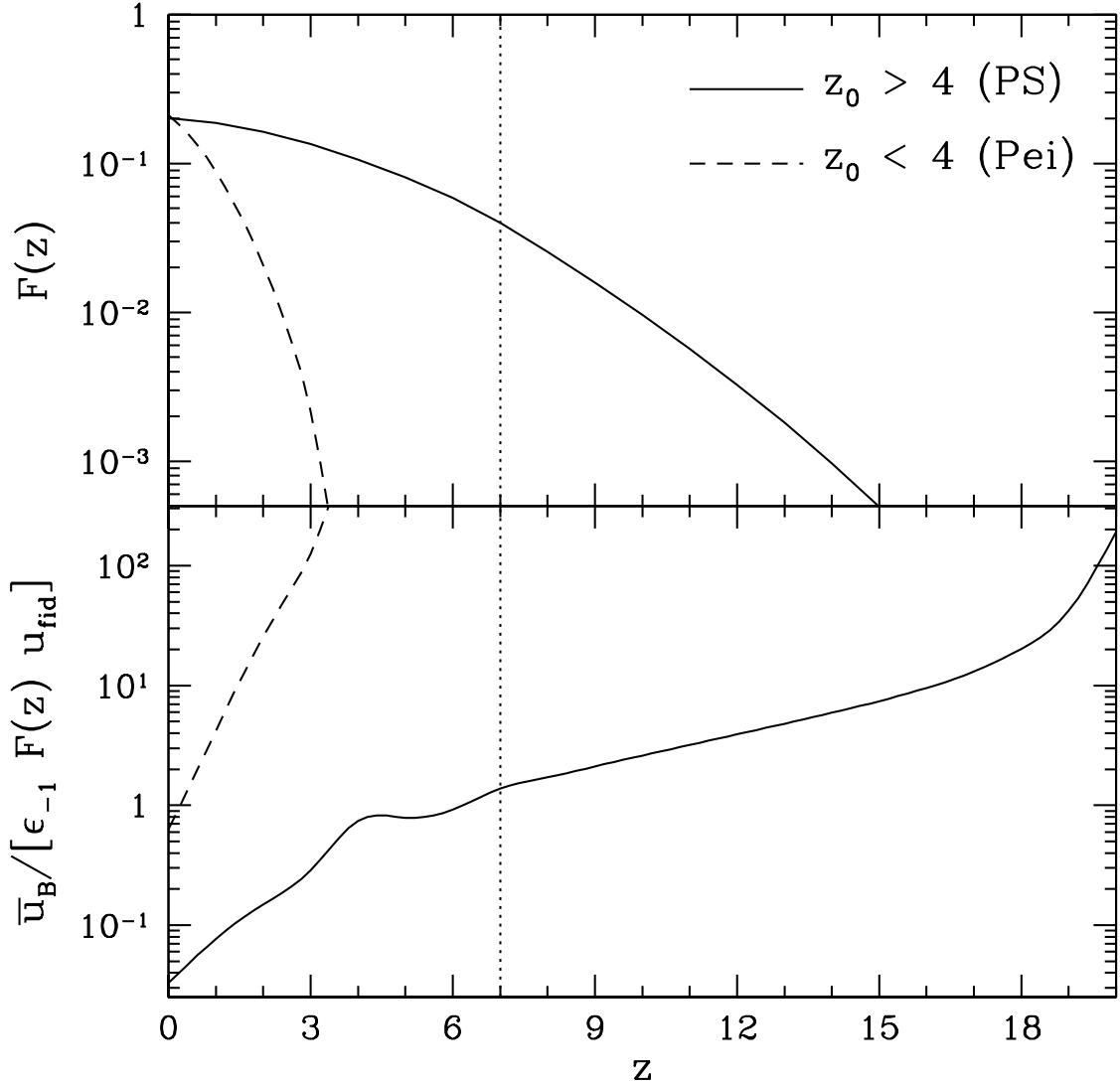}
\caption{\emph{Upper Panel:} Volume filling fraction of magnetized bubbles
$F(z)$, as a function of redshift, for the RLQ model.  \emph{Lower
Panel:} Ratio of magnetic 
energy per filled volume, $\bar{u}_B/[\epsilon_{-1} F(z)]$, to the
fiducial thermal energy density $u_{fid} =
3 n(z) k T$, where $T = 10^4 \kel$, as a function of redshift.  The solid
lines indicate the contribution from high-redshift quasars calculated using
the Press-Schechter formalism for $z_0 > 4$.  The dashed lines indicate
the contribution determined from the Pei (1995) luminosity function 
for $z_0 < 4$.  All curves assume $f_R^{PS} = 0.01$, $f_R^{Pei} =
0.1$, $\tau_q=10^7 \yr$, and $f_d=0.8$. The
minimum halo mass $M_{h,min}$ is determined by atomic cooling before
reionization and infall suppression afterward.  The vertical dotted line
indicates the assumed redshift of reionization, $z_r = 7$.  }
\end{figure}

\begin{figure} [t]
\plotone{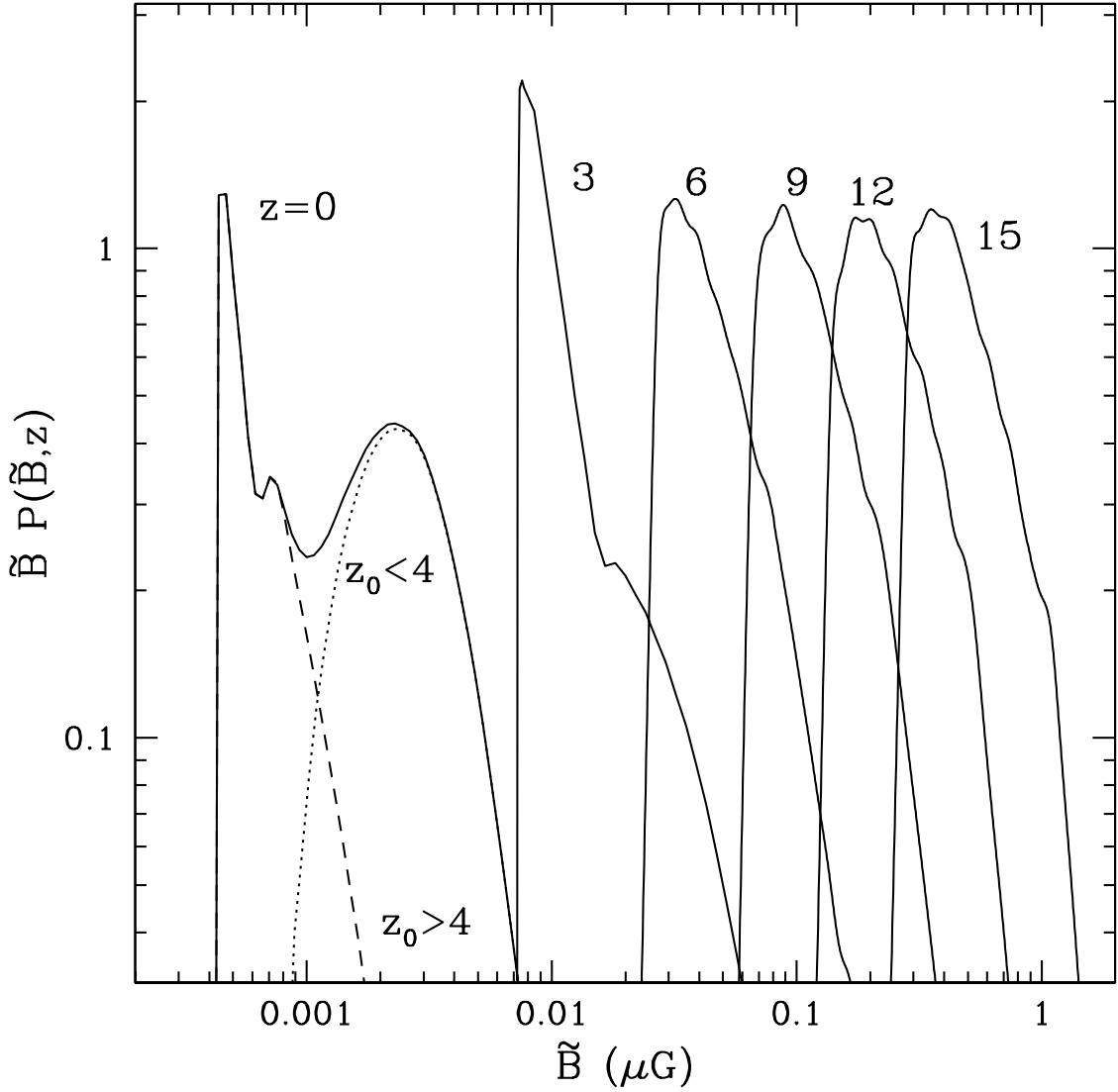}
\caption{ Probability distribution of bubble magnetic field,
$P(\tilde{B},z)$, for $z=0,3,6,9,12,$ and $15$, from left to
right, where $\tilde{B} = B/(\sqrt{\epsilon_B/0.1})$. 
The dotted curve shows the contribution to the $z=0$ distribution function
from quasars at $z_0 < 4$ and the dashed curve shows the contribution from
quasars at $z_0 > 4$.  The calculation assumes $f_R^{PS} = 0.01$, $f_R^{Pei} =
0.1$, $f_d = 0.8$, $\tau_q=10^7 \yr$, $\epsilon_B = 0.1$, and $z_r = 7$.  }
\end{figure}

\begin{figure} [t]
\plotone{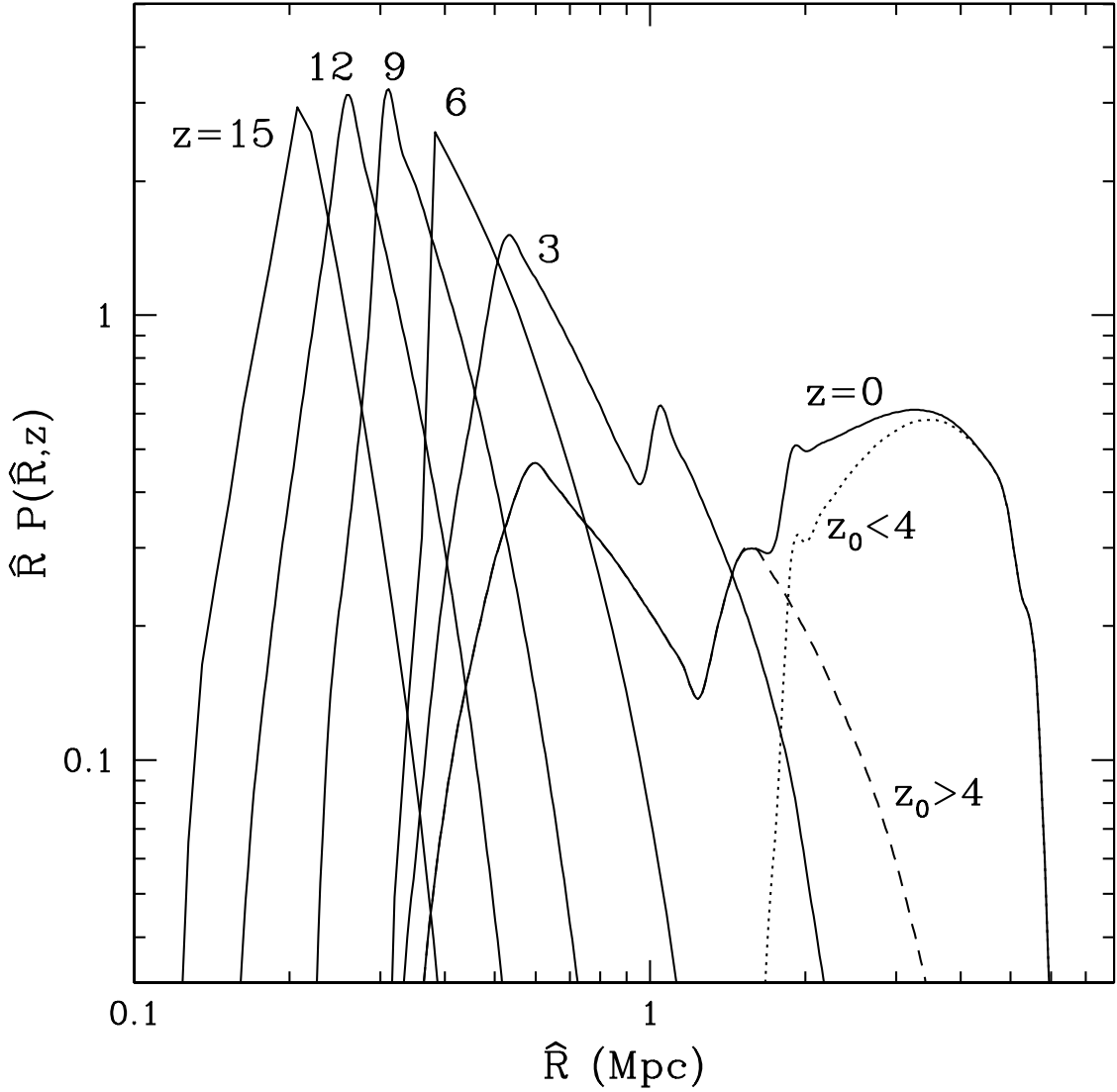}
\caption{Probability distribution of bubble comoving
radius $P(\hat{R},z)$, for $z=15,12,9,6,3,$ and $0$, from left to right.
The dotted curve shows the contribution to the $z=0$ distribution function
from quasars at $z_0 < 4$ and the dashed curve shows the contribution from
quasars at $z_0 > 4$.  The calculation assumes $f_R^{PS} = 0.01$, $f_R^{Pei} =
0.1$, $f_d = 0.8$, $\tau_q=10^7 \yr$, $\epsilon_B=0.1$, and $z_r = 7$.  }
\end{figure}

\begin{figure} [t]
\plotone{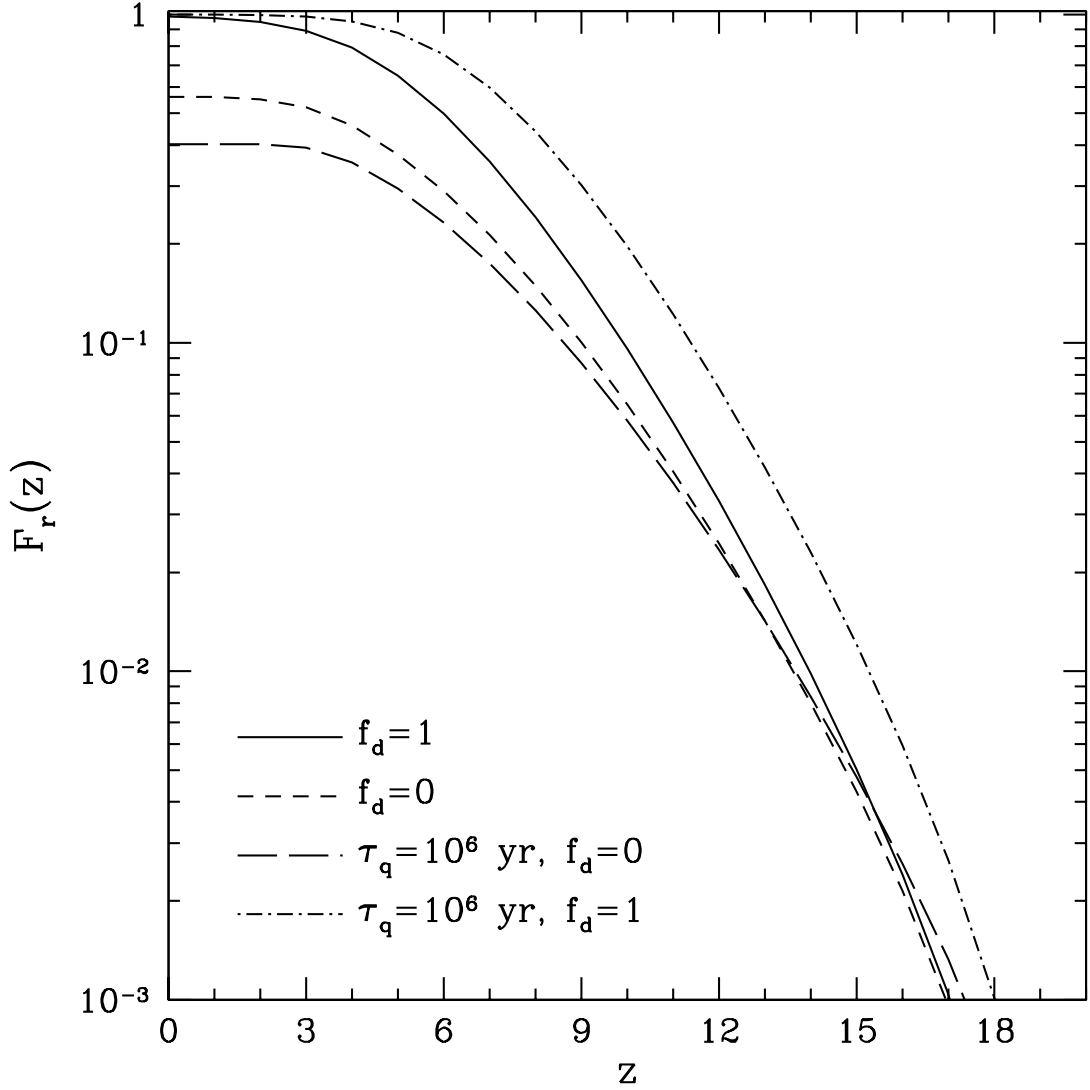}
\caption{Space filling fraction of ionized material, $F_r(z)$, as a
function of redshift, for several different scenarios.  The quasar
population with $z_0 > 4$ is included.  Results are shown for $f_d=1$
(solid line) and $f_d=0$ (short dashed line); each of these
cases has $f_{BAL}^{PS} = 0.1$, $\tau_q=10^7 \yr$, and $\epsilon_B = 0.1$.
Also shown are cases in which $f_{BAL}^{PS} = 1.0$, $\tau_q=10^6 \yr$, and
$\epsilon_B = 0.1$, with $f_d=1$ (long dashed line) and $f_d=0$ (dot-dashed
line).  The minimum halo mass is determined by the requirement of efficient
atomic cooling at all redshifts.  }
\end{figure}

\end{document}